# Intertwined Charge Stripes and Majorana Zero Modes in An Iron-Based Superconductor

Yu Liu[1,†], Li-Xuan Wei[1,†], Qiang-Jun Cheng[1,†], Zhenhua Zhu[1], Xin-Yu Shi[1], Cong-Cong Lou[1], Yong-Wei Wang[1], Ze-Xian Deng[1], Ming-Qiang Ren[1,2], Dong E. Liu[1], Ziqiang Wang[3,*], Xu-Cun Ma[1,4,*], Jin-Feng Jia[2,5,6], Qi-Kun Xue[1,2,4,6,7,*], Can-Li Song[1,4,*]

*[1]Department of Physics and State Key Laboratory of Low-Dimensional Quantum Physics, Tsinghua University, Beijing 100084, China*

*[2]Shenzhen Institute for Quantum Science and Engineering and Department of Physics, Southern University of Science and Technology, Shenzhen 518055, China*

*[3]Department of Physics, Boston College, Chestnut Hill, Massachusetts 02467, USA*

*[4]Frontier Science Center for Quantum Information, Beijing 100084, China*

*[5]Key Laboratory of Artificial Structures and Quantum Control (Ministry of Education), School of Physics and Astronomy, Shanghai Jiao Tong University, Shanghai 200240, China*

*[6]Hefei National Laboratory, Hefei 230088, China*

*[7]Beijing Academy of Quantum Information Sciences, Beijing 100193, China*

In type-II superconductors, magnetic fields modulate the amplitude and phase of the superconducting order parameter, forming quantized vortices where superconductivity is locally suppressed and exotic bound states or competing electronic orders emerge. Using spectroscopic-imaging scanning tunneling microscopy on epitaxial $Ba(Fe_{0.94}Co_{0.06})_2As_2$ films, we discover an incommensurate charge-stripe order aligned with the Fe-Fe bond direction. These charge modulations become strongly enhanced at the vortex core and extend far into the vortex halo. Strikingly, the charge order modulates Andreev bound states of vortices at non-zero energies, producing abelian vortices with half-odd-integer level quantization and non-abelian vortices with integer-quantized core states that host a Majorana zero mode. The distinct vortex types are distinguished by the registry of their centers relative to the charge-stripe pattern and remain robust in ultrathin (2.5-unit-cell) films. Our findings reveal a density-wave-textured vortex matter and provide fresh insights into intertwined phenomena in iron-based superconductors.

[†]These authors contributed equally to this work

*Corresponding authors. Emails: clsong07@mail.tsinghua.edu.cn, ziqiang.wang@bc.edu, xucunma@mail.tsinghua.edu.cn, qkxue@mail.tsinghua.edu.cn

Iron-based superconductors (IBSs) have, since their discovery, emerged as a versatile material platform for exploring unconventional superconductivity, strong correlations, and symmetry-breaking electronic states such as nematicity and collinear antiferromagnetism [1-3]. These correlated electronic orders frequently coexist with the superconducting (SC) condensation and become more prominent inside magnetic vortices, where superconductivity is locally weakened. Rotational symmetry breaking electronic nematicity [2-6], for example, has been visualized as elongated vortex cores in FeSe and $RbFe_2As_2$ [7-9]. By contrast, the translational symmetry breaking charge order – long established in underdoped cuprates [10-13] and recently attracting widespread interest across twisted moiré materials [14], transition metal dichalcogenides [15,16], kagome metals [17], and nickelate superconductors [18] – has remained elusive in IBSs [19,20], the second major high-temperature ($T_c$) family. This absence constitutes one of the unresolved gaps in the comparative phase diagrams of unconventional superconductors and has hindered a unified understanding of how a charge order integrates into the hierarchy of correlated electronic states and shapes the low-energy excitations in these prominent quantum materials.

The IBS platform is further enriched by strong spin-orbit coupling (SOC), leading to a $Z_2$ nontrivial band structure with a helical Dirac-fermion topological surface state (TSS) [21-23]. In some situations, the TSS coexists with a three-dimensional (3D) topological Dirac semimetal (TDS) protected by $C_4$ rotational symmetry [24,25]. The magnetic vortices of these SC TSS can host Majorana zero modes (MZMs) in a single-material realization of the Fu-Kane proposal [26], obviating the need for hybrid heterostructures [27-29]. As non-abelian anyons, MZMs are promising building blocks for fault-tolerant quantum computation. Moreover, IBSs lie in the quantum limit where the SC gap $\Delta$ is comparable to the Fermi energy $E_F$ [30], such that vortex-induced Caroli-de Gennes-Matricon (CdGM) bound states appear as discrete levels with spacing $\delta_E \sim \Delta^2/E_F$ [31]. Zero-bias conductance peaks – candidate MZMs associated with integer-quantized CdGM states – have been increasingly observed by spectroscopic-imaging scanning tunneling spectroscopy (SI-STM) in several IBS families, including iron chalcogenides [32-37], $CaKFe_4As_4$ [38], and LiFeAs [19]. However, such features appear only in a subset of vortices, termed Majorana or non-abelian vortices. Their coexistence with conventional abelian vortices, showing half-odd-integer CdGM levels and no zero-bias peak, in the same system remains a subject of debate [39-43]. Here we advance these frontiers of correlation and topology by revealing a unidirectional charge-density-wave (CDW) order within magnetic vortices of $Ba(Fe_{0.94}Co_{0.06})_2As_2$ films and its unusual modulations on integer- and half-odd-integer-quantized CdGM states – two phenomena intrinsically intertwined, as neither can exist without the other.

To obtain high-quality IBSs suitable for resolving competing order and vortex-core excitations with high resolution by SI-STM, we have employed molecular beam epitaxy (MBE) to grow archetypal ferro-pnictide $Ba(Fe_{0.94}Co_{0.06})_2As_2$ (BFCA) thin films near optimal doping on $SrTiO_3(001)$ [Figs. 1(a) and S1], as described in Method [44]. The SC transition temperature $T_c$ (~ 30 K) exceeds the bulk value (~ 22 K) [45], indicating superior crystallinity and reduced disorder in BFCA films [46]. Figure 1(b) displays spatially resolved differential conductance spectra, $g(\boldsymbol{r}, V) \equiv \mathrm{d}I/\mathrm{d}V(\boldsymbol{r}, V)$, measured on a 10-UC-thick film, which reflect the local density of states (DOSs) as a function of energy $E = eV$ (with $e$ the elementary charge) at coordinate $\boldsymbol{r}$. These spectra reveal two fully opened superconducting gaps, $\Delta_l \approx 4.9$ meV and $\Delta_s \approx 2.8$ meV, attributable to the multiband character of BFCA [47], with the DOSs fully suppressed over ± 2.5 meV. This demonstrates the substantial elimination of disorder-induced subgap states common in cleaved BFCA crystals [48]. Furthermore, the films contain very few impurities, with the impurity-induced subgap states well away from $E_F$. These attributes enable the discovery of incommensurate charge-stripe-textured vortices and MZMs in BFCA under perpendicular magnetic fields $B$ [Fig. 1(a)].

Upon applying a magnetic field, unidirectional DOS modulations develop within every vortex core and extend outward in BFCA, as revealed by energy-dependent conductance maps $g(\boldsymbol{r}, E)$ around a representative vortex #1 [Figs. 1(c) and S2]. These stripes consist of uniaxial oscillations of vortex-induced states along the orthorhombic $a$ axis – the antiferromagnetic Fe-Fe bond direction, with no discernible modulations along the orthogonal $b$ axis. Fourier amplitudes $g(\boldsymbol{q}, E)$, derived from $g(\boldsymbol{r}, E)$, show that the modulations are confined exclusively to the $q_a$ direction [Fig. 1(d)]. This pronounced anisotropy aligns with the vortex-core elongation in Figs. 1(a) and 1(c). In contrast to the nematicity-induced vortex elongation reported in FeSe [7,8], vortex-core states in BFCA display translational-symmetry-breaking modulations that propagate far beyond the core boundary, characterized by the coherence length $\xi \approx 2.5$ nm (Supplemental Material, Sec. 1, Fig. S3), into the vortex halos along the $a$ direction. Despite the fully opened SC gap at zero field [Fig. 1(b)], the electronic spectrum inside the vortex cores is gapless, filled with an underlying continuum of states [Fig. 1(e)]. The striped modulations become significantly suppressed at zero energy and at energies beyond ± 3.0 meV [Fig. S2], while weak modulations with the same wavevector remain detectable outside the superconducting gaps [Fig. S4]. Remarkably, the $g(\boldsymbol{r}, E)$ modulations are invariably in-phase either above or below $E_F$, but undergo a π-phase shift upon reversal of the sample-bias polarity [Figs. 1(c) and S2]: brighter stripes in the occupied states ($E < 0$, charge accumulation) are darker in the empty states ($E > 0$), and vice versa (charge depletion), a hallmark of unidirectional CDW. Such observations fundamentally differ from the behavior expected for conventional Friedel-like oscillations of CdGM bound states, in which neighboring vortex-bound levels exhibit anti-phase spatial modulations [30, 31]. This indicates that the stripe signal is not simply a discrete in-gap CdGM interference pattern, consistent with its observability at energies outside the superconducting gap. Instead, these results suggest that the vortex-core electronic structure is influenced by the underlying charge-stripe modulation.

We further confirm these observations and quantify the charge-stripe wavelength $\lambda$ through measuring high-resolution line-cut spectroscopy along the $a$ direction across the vortex center $r = 0$ [Fig. S5]. Beyond the decay away from $r = 0$ and the universal contrast reversal, pronounced spatial modulations of $g(\boldsymbol{r}, E)$ persist well within $\pm\Delta_s$ except for $E \sim 0$, as more clearly revealed by their Fourier transforms [Fig. 1(f)]. The modulation wavevector is $Q \approx 0.47$ Å$^{-1}$, corresponding to $\lambda \sim 1.33$ nm $\approx 4.7a_{Fe}$, where $a_{Fe}$ represents the Fe-Fe bond length, incommensurate with the underlying lattice. This short $\lambda$, roughly half of $\xi$, together with the elongated vortex cores, enables direct visualization of charge stripes in BFCA films. This incommensurate charge-stripe order competes with superconductivity and becomes strongly enhanced as the condensate is locally weakened inside vortices. It is therefore distinct from the apparent coexistence of superconductivity and strain-induced charge stripes in LiFeAs [19], where lattice wrinkles dominate and vortices play little role.

The charge modulations are consistently observed around every vortex core (e.g., vortex #2 in Fig. S6), independent of film thickness. Importantly, all vortex cores host multiple discrete, particle-hole symmetric CdGM states, superimposed and broadened by the continuum background of states [Figs. 1(e) and 2(a)]. This behavior differs sharply from previous observations in bulk BFCA crystal and its hole-doped counterpart [48,49]. Strikingly, all discrete CdGM levels, except the one precisely at $E = 0$, display $a$-axis modulations [Figs. 2(b) and S5]. This is corroborated by spatially resolved $g(\boldsymbol{r}, E)$ profiles along the $a$ direction across vortex #2 [Fig. 2(c)], where the exponential decay from $r = 0$ is periodically modulated with wavelength $\lambda$, accompanied by the charge-stripe-encoded contrast reversal between occupied and empty states. Again, no modulation is detectable along the $b$ axis within our measurement accuracy [Fig. S7].

In the quantum limit, vortex-induced Andreev bound states, namely the CdGM states are quantized into discrete energy levels $E_\mu = \mu\Delta^2/E_F$, where $\mu = \pm 1/2, \pm 3/2, \ldots$ is a half-odd-integer angular quantum number [31]. For vortices from the SC TSS (see inset in Fig. 2(a)), an additional Berry phase from the helical Dirac-fermions shifts the quantization to integer values $\mu = 0, \pm 1, \pm 2, \ldots$[26,50,51]. The discrete states observed in Figs. 1(e) and 2(a) are nicely consistent with the integer-quantized sequence of the CdGM states, with the zero-energy level at $\mu = 0$ corresponding to the non-abelian MZM. These discrete CdGM states, including the MZM, are remarkably robust, showing no energy splitting away from the vortex center [Fig. S7] and under variations in tunneling barrier.

Extracting the discrete $E_\mu$ from either fitted d$I$/d$V$ spectra [Fig. 2(a)] within vortex #2 or from numerical analysis [Fig. S7] yields a large collection of $E_\mu$ values, whose histogram reveals a uniformly spaced, integer-quantized sequence [Fig. 2(d)]. The mean energy $E_\mu$ scales linearly with integer index $\mu$ [Fig. 2(e)], passing through zero and giving a level spacing $\delta_E = \Delta^2/E_F \sim 0.49$ meV, supporting the identification of an ensemble of Majorana vortices. The observation of integer-quantized $E_\mu$ levels up to $\mu = \pm 5$ and the zero-energy MZM provide compelling spectroscopic evidence for SC TSS in BFCA films [50,51], placing the system in both the clean and quantum limits [31]. Notably, while the non-zero CdGM states couple strongly to the charge modulations, the MZM exhibits negligible spatial modulation, attesting to its charge neutrality. Furthermore, the MZM is robust against impurities and persists even in 2.5-UC-thick (~ 3.3 nm) BFCA films [Fig. 3(a)] – the thinnest samples with atomically flat terraces suitable for STM imaging of vortices. This thickness limit is comparable to the previous record (~ 4.1 nm) for hosting TSS and MZM in $Bi_2Te_3/NbSe_2$ heterostructures [28].

Interestingly, a subset of vortices exhibits discrete CdGM states that follow the conventional half-odd-integer quantization sequence, i.e., $E_\mu = \mu\Delta^2/E_F$ with $\mu = \pm 1/2, \pm 3/2, \ldots$, and lack MZMs [Figs. 3(b) and S9]. This dramatic change occurs despite similar charge-stripe-induced unidirectional modulations inside the vortex halos [Fig. 3(b)], including contrast reversals between positive and negative $\mu$ states [Fig. S9(b)]. To assess universality and exclude vortex-specific artifacts, we systematically studied vortices across epitaxial BFCA films of varying thicknesses [Fig. S10]. Histograms of $E_\mu$, normalized by $\Delta^2/E_F$, reveal that the discrete vortex-core states follow integer (Majorana vortices) or half-odd-integer (conventional vortices) quantization sequences [Fig. 3(c)], independent of film thickness and $\delta_E$. In both cases, finite-energy CdGM states exhibit robust unidirectional charge modulations. These findings, established with statistical significance, reveal that the spectroscopic dichotomy is an intrinsic property of charge-stripe modulated vortices.

To gain further insight into charge-vortex coupling and its potential role in shaping the distinct Majorana and conventional vortices, we analyzed sixty vortices using high-resolution line-cut spectra and characterized each by three key attributes (Supplemental Material, Sec. 2): the charge-stripe wavelength $\lambda$, the energy spacing $\delta_E$, and the distance $d$ (in units of $\lambda$) between the vortex center to the nearest rising node ($g(\boldsymbol{r}, E < 0) = 0$ and $dg(\boldsymbol{r}, E < 0)/dr > 0$) of the charge modulations along the $a$ axis – a quantitative measure of the vortex registry relative to the charge configuration. Figure 4(a) presents all Majorana (orange) and conventional (gray) vortices in this 3D parameter space. The charge stripes are remarkably stable with $\lambda \approx 1.30 \pm 0.07$ ($Q \approx 0.48 \pm 0.03$ Å$^{-1}$), whereas $\delta_E$ varies nearly fourfold (0.4 ~ 1.6 meV), reflecting a high manipulability of the vortex-core states. However, $\delta_E$ does not distinguish Majorana from conventional vortices, which is difficult to attribute their dichotomy to TSS and bulk-band differences [34].

Interestingly, the vortex dichotomy correlates with $d$ – the registry of vortex center relative to the charge stripes. Majorana vortices preferentially nucleate near charge nodes, whereas conventional vortices are more frequently found near the antinodes, i.e., the extrema – either maxima or minima – of the charge stripes. A small fraction (<5%) of vortices deviates from this trend, which probably originates from finite measurement precision (Supplementary Material, Sec. 2). This finding suggests that the displacement of the vortex center to the charge nodal lines regulates the CdGM quantization sequence and control the emergence of MZMs. The lacking of vortex pinning to specific CDW sites is likely due to the charge-stripe wavelength $\lambda$ being comparable to the vortex-core dimension $\xi \sim 2.5$ nm. Consistent with this picture, nanometer-scale displacement of vortex cores under repeated field applications can switch the vortex between Majorana and conventional character (Supplemental Material, Sec. 3 and Fig. S11).

A local modulation of the chemical potential by the charge stripes – strong at the antinodes and weak at the nodes – cannot, at least in its simple form, explain the vortex dichotomy (Supplemental Material, Sec. 4), as no correlation is found between the vortex type and the $E_F$-related CdGM level spacing $\delta_E = \Delta^2/E_F$ [Fig. 4(a)]. In contrast, spatially modulated superconducting order parameter $\Delta(\boldsymbol{r})$ has been discussed in systems where superconductivity coexists with CDW order [52,53], as previously revealed in $NbSe_2$ [54]. Similar spectroscopic signatures, such as modulations of the superconducting gap and coherence-peak intensity, are also observed both outside vortex cores [Figs. 2(b) and S12] and in the Meissner state of BFCA (Supplemental

Material, Sec. 5 and Fig. S13). These modulations share the same wavevector $Q$ as the CDW, which is nearly twice that of $EuRbFe_4As_4$ [20]. Based on our theoretical analysis and phenomenological model calculations (Supplemental Material, Sec. 6 and Sec. 7) [55-57], we show that when a vortex nucleates at nodes of this unidirectional superconductivity modulation $\Delta_Q(x)$ ($x$ is the coordinate along the $a$ axis, Fig. 4(b)), the mirror-odd symmetric $\Delta_Q(x)$ with $x = 0$ matches the vortex order parameter $|\Delta(r)|e^{i\theta}$ in symmetry ($r$ and $\theta$ are the polar coordinates), preserving the integer-quantized CdGM bound states and MZMs [Figs. S14(a) and (b)]. Conversely, when the vortex center aligns with the antinodes of the $\Delta_Q(x)$ [Fig. 4(c)], the resulting symmetry mismatch can induce phase dislocations [58-62], introducing phase discontinuities that destabilize the integer quantization sequence and remove the zero-energy modes [Fig. S14(c)]. While this analysis captures the essential physics and offers a viable theoretical framework, both charge and spatial superconducting modulations are likely intertwined in shaping the vortex-core spectrum. A complete understanding of the coupled influence, as well as the dual character of charge- and $\Delta_Q(x)$-modulated vortices, await future self-consistent theoretical calculations.

We stress that the discovery of charge stripes resolves a long-standing missing component in the phase diagram of IBSs. Moreover, the enhancement of the charge stripes inside vortex cores provides an ideal platform for probing how charge ordering couples and controls low-energy vortex core excitations. In particular, the observation of density-wave-textured magnetic vortices transcends the conventional understanding of vortex matter and renders a single-particle description inadequate [31]. In the standard picture, discrete CdGM states $E_\mu$ were expected to exhibit spatial intensity oscillations with a wavelength $\lambda_F = 2\pi/k_F$, where $k_F$ is the Fermi wavevector and closely related with $\delta_E = \Delta^2/E_F$ [31,63,64]. These oscillations typically form concentric rings, with adjacent $E_\mu$ states modulating approximately out of phase [30,38]. Our results deviate fundamentally from this picture: vortex-core states exhibit modulations at a fixed wavelength $\lambda$ corresponding to the charge-stripe order across all non-zero energies, with pronounced contrast reversals between $\pm E_\mu$ states but no change in period despite substantial variations in $\delta_E$ [Fig. 4(a)]. These findings provide strong evidence for a pre-existing electronic modulation inside magnetic vortices, attesting to a new form of correlated vortex matter in which vorticity and density-wave orders intertwine to manipulate the vortex-core spectrum and the stability of the MZM. Notably, the wavevector $Q = 2\pi/\lambda$ nearly matches twice the Fermi vector $k_F \approx 0.22$ Å$^{-1}$ of the $k_z$-independent $\eta$ band near the M point of the Brillouin zone [47,65], suggesting a possible link between the observed charge-stripe order and electronic correlations of the $\eta$ band, potentially enhanced by intertwined nematic and antiferromagnetic fluctuations [66]. An even more intriguing possibility is that the charge-stripe order may originate from the TDS band self-doped away from its Dirac point. Distinguishing these two possibilities would require further momentum-resolved measurements, such as high-resolution angle-resolved photoemission spectroscopy or quasiparticle-interference imaging, to directly determine whether the CDW gap develops predominantly on the $\eta$ band or on the TDS band. Notably, if the CDW couples strongly to the TDS states, it could reconstruct the topological electronic structure by breaking the $C_4$ rotational symmetry and gapping out the TDS states that would otherwise hybridize with TSS, thereby destroying the MZM [19,43] and producing gapless vortex excitations [49].

Our unprecedented observation of density-wave-textured vortices carry profound implications extending beyond the microscopic mechanisms of charge-stripe order and its coupling to vortex-induced Andreev bound states in BFCA. Like the checkerboard charge order in cuprates [10-13], the charge-stripe order in BFCA becomes significantly enhanced inside magnetic vortices where superconductivity is locally suppressed. It is therefore tempting to speculate that complete quenching of superconductivity – such as under higher magnetic fields – may stabilize a macroscopic charge-stripe-ordered ground state. Alternatively, fluctuating charge stripes may represent key low-energy excitations in the homogeneous SC phase, potentially intertwined with the orbital and spin fluctuations that characterize the normal state [67-69]. These insights provide a new perspective on the interplay between electronic correlation and superconductivity in unconventional high-$T_c$ materials. Moreover, we have shown that the charge-stipe modulations can couple to and regulate the Andreev bound states between integer- and half-odd-integer quantization sequences inside vortex cores. Beyond establishing an encouraging route toward realizing and controlling MZMs in charge-ordered high-$T_c$ superconductors, our findings offer a versatile and much needed platform for probing the fundamental interplay among the intertwined electronic orders and vorticity. Finally, ultrathin IBS films down to only 2.5 UCs host robust MZMs via charge stripe-vortex registry tuning, opening a pathway toward scalable vortex-Majorana architectures, including patterned vortex pinning and electrostatic gating. This constitutes, to our knowledge, the first observation of an MZM-consistent zero mode in lithography-compatible high-$T_c$ superconductor films, bridging fundamental topological superconductivity with its technological realization toward fault-tolerant quantum computation.

**Acknowledgments**

The work was financially supported by the Natural Science Foundation of China (Grant No. 12474130, Grant No. 12134008, Grant No. 12141403 and Grant No. 52388201), the National Key Research and Development Program of China (Grant No. 2022YFA1403100). Z.W. is supported by the U.S. Department of Energy, Basic Energy Sciences Grant DE-FG02-99ER45747.

Y. Liu, L. X. Wei and Q. J. Cheng contributed equally to this work

***Data availability*** The data that support the findings of this article are openly available [70].

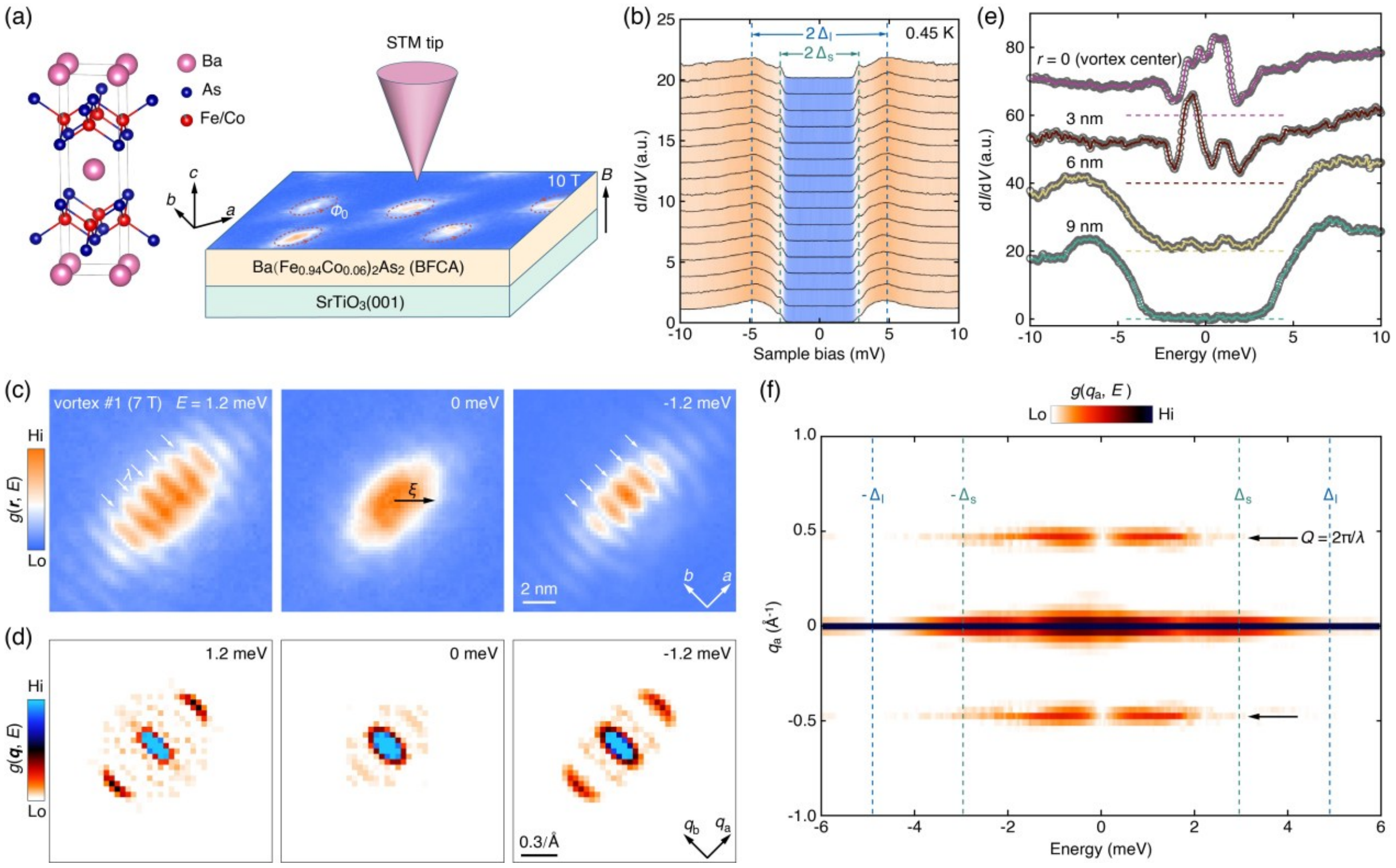

**Figure 1** (a) Crystal structure of tetragonal BFCA (left) and schematic STM setup (right) for visualizing Abrikosov vortices (dashed ellipses) in BFCA films prepared on $SrTiO_3$. The *a* and *b* axes are oriented with the Fe-Fe bond directions, while the *c* axis aligns with the stacking direction of the FeAs layers. A magnetic field *B* is applied perpendicular to the sample surface, with vortex cores elongated along the antiferromagnetic *a* axis. (b) d*I*/d*V* spectra taken along a 5-nm trajectory at 0.45 K, revealing two distinct superconducting gaps, marked by green ($\Delta_s$) and blue ($\Delta_l$) dashes, respectively. Setpoint: $V = 10$ mV, $I = 0.5$ nA. (c) Energy-resolved $g(\boldsymbol{r}, E)$ maps measured over a 14.5 nm × 14.5 nm field of view in a 10-UC-thick film, revealing unidirectional charge modulations (marked by the white arrows) within the halo of magnetic vortex #1. For clarity, the maps are mirror-symmetrized relative to the *a* axis through the vortex center. Setpoint: $V = 10$ mV, $I = 0.12$ nA. (d) Amplitude $g(\boldsymbol{q}, E)$ obtained via Fourier transforms of $g(\boldsymbol{r}, E)$. (e) Representative d*I*/d*V* spectra acquired around vortex #1, with *r* representing the distance away from the vortex center ($r = 0$). The vortex center is determined as the coordinate where $g(\boldsymbol{r}, E = 0)$ reaches its maximum along the *a* axis throughout. The spectra are vertically offset for clarity, and their zero-conductance baselines are indicated by horizontal dashes of corresponding colors. Setpoint: $V = 10$ mV, $I = 0.15$ nA. (f) Color plot of $g(q_a, E)$ extracted from Fourier transforms of $g(\boldsymbol{r}, E)$ profiles, taken across vortex #1 along the *a* axis at various energies. Black arrows indicate a non-dispersive scattering vector at $Q \approx 0.47$ Å$^{-1}$ that arises from vortex-induced charge modulations and predominantly exists at non-zero energies within $\pm\Delta_l$.

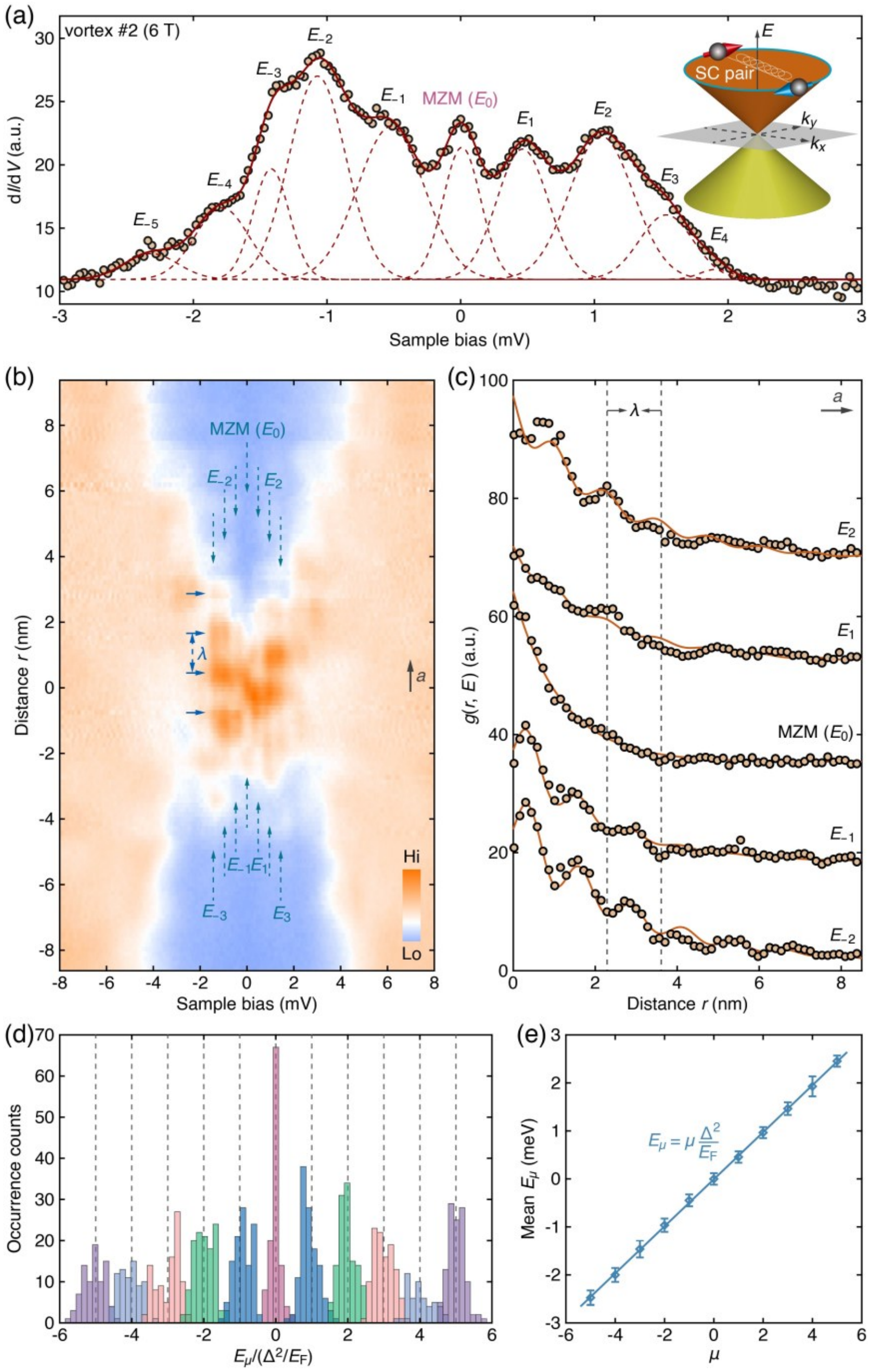

**Figure 2** (a) d$I$/d$V$ spectrum (setpoint: $V$ = 6 mV, $I$ = 0.2 nA) measured at the center of vortex #2, displaying a zero-bias conductance peak attributed to MZM and discrete bound states at $E_\mu$ ($\mu$ = integers) as labelled. The solid curve represents a multi-Gaussian fit to the experimental data (orange circles), with dashed lines denoting individual Gaussian components. (b) Intensity plot of line-cut d$I$/d$V$ spectra across the center ($r$ = 0) of Majorana vortex #2 along the $a$ axis. Green and black arrows indicate discrete CdGM states and charge modulations, respectively. Setpoint: $V$ = 10 mV, $I$ = 0.15 nA. (c) Spatially resolved $g(\boldsymbol{r}, E)$ profiles extracted from (b). Discrete CdGM states at finite $E_\mu$ exhibit spatial modulations with a wavelength $\lambda \approx 4.7a_{\mathrm{Fe}}$ and π-phase shift across $E_F$, guided by vertical dashes. Solid lines represent fits of the $g(\boldsymbol{r}, E)$ profiles to either exponential ($E_0$) or oscillatory decay ($E_{\pm1}$ and $E_{\pm2}$) functions. (d) Histogram of $E_\mu$ values extracted from 256 d$I$/d$V$ spectra within vortex core #2, normalized by its discrete energy level spacing $\Delta^2/E_F$. (e) Linear fit of mean $E_\mu$ as a function of the integer level index $\mu$, yielding $\Delta^2/E_F \sim 0.49$ meV between adjacent CdGM levels.

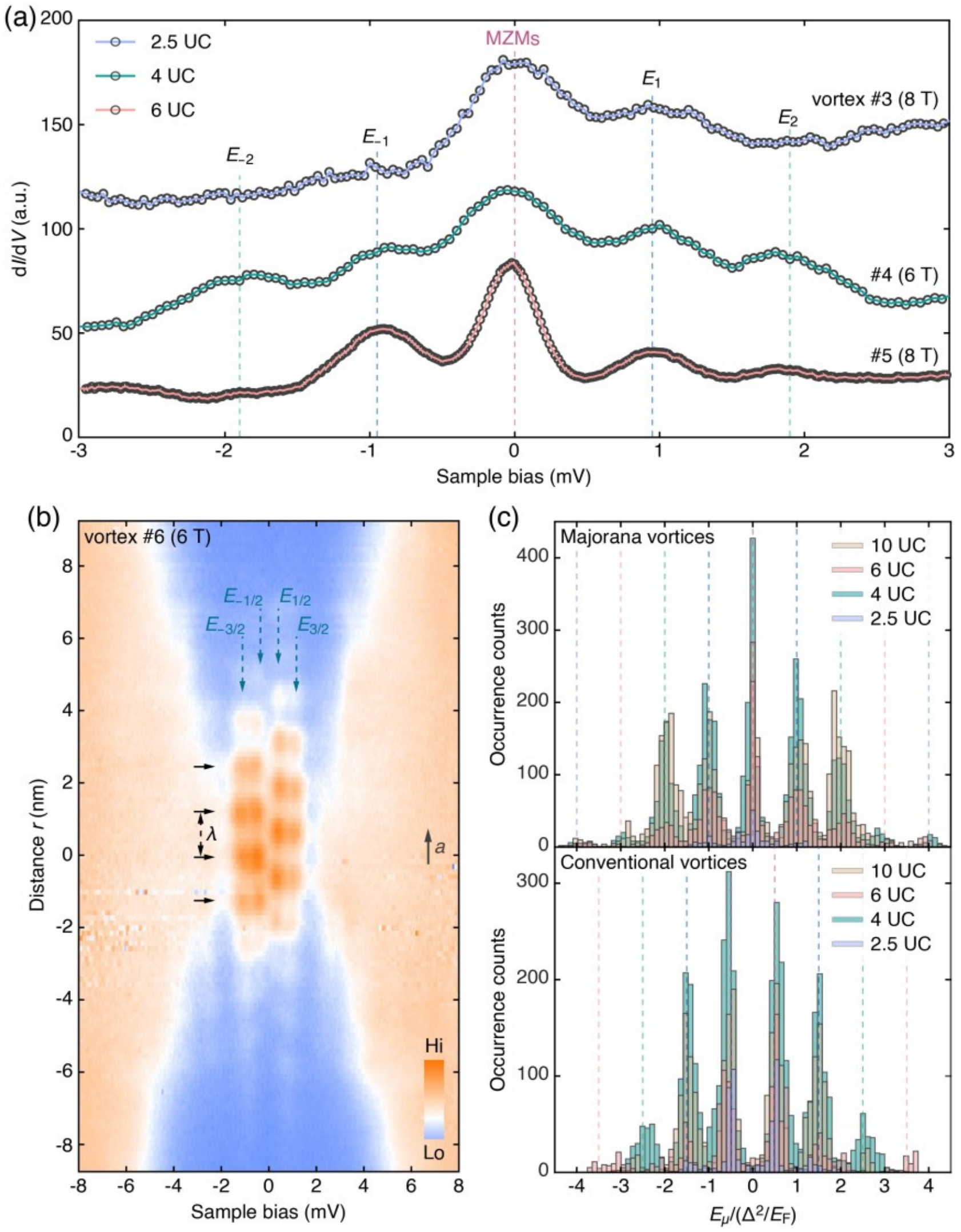

**Figure 3** (a) d$I$/d$V$ spectra measured near the centers of Majorana vortices #3 (8 T), #4 (6 T) and #5 (8 T) in BFCA films of varying thicknesses, showing integer-quantized CdGM states and MZMs. Setpoint: $V = 8$ mV, $I = 0.25$ nA. (b) Intensity plot of line-cut d$I$/d$V$ spectra across a conventional vortex # 6 in 10-UC-thick BFCA films, taken along the $a$ axis. Setpoint: $V = 8$ mV, $I = 0.25$ nA. (c) Histograms of normalized $E_\mu$, extracted from > 11000 $dI/dV$ spectra around 253 vortices, revealing two distinct quantization sequences: integer levels (top panel) and half-odd-integer levels (bottom panel), independent of film thickness.

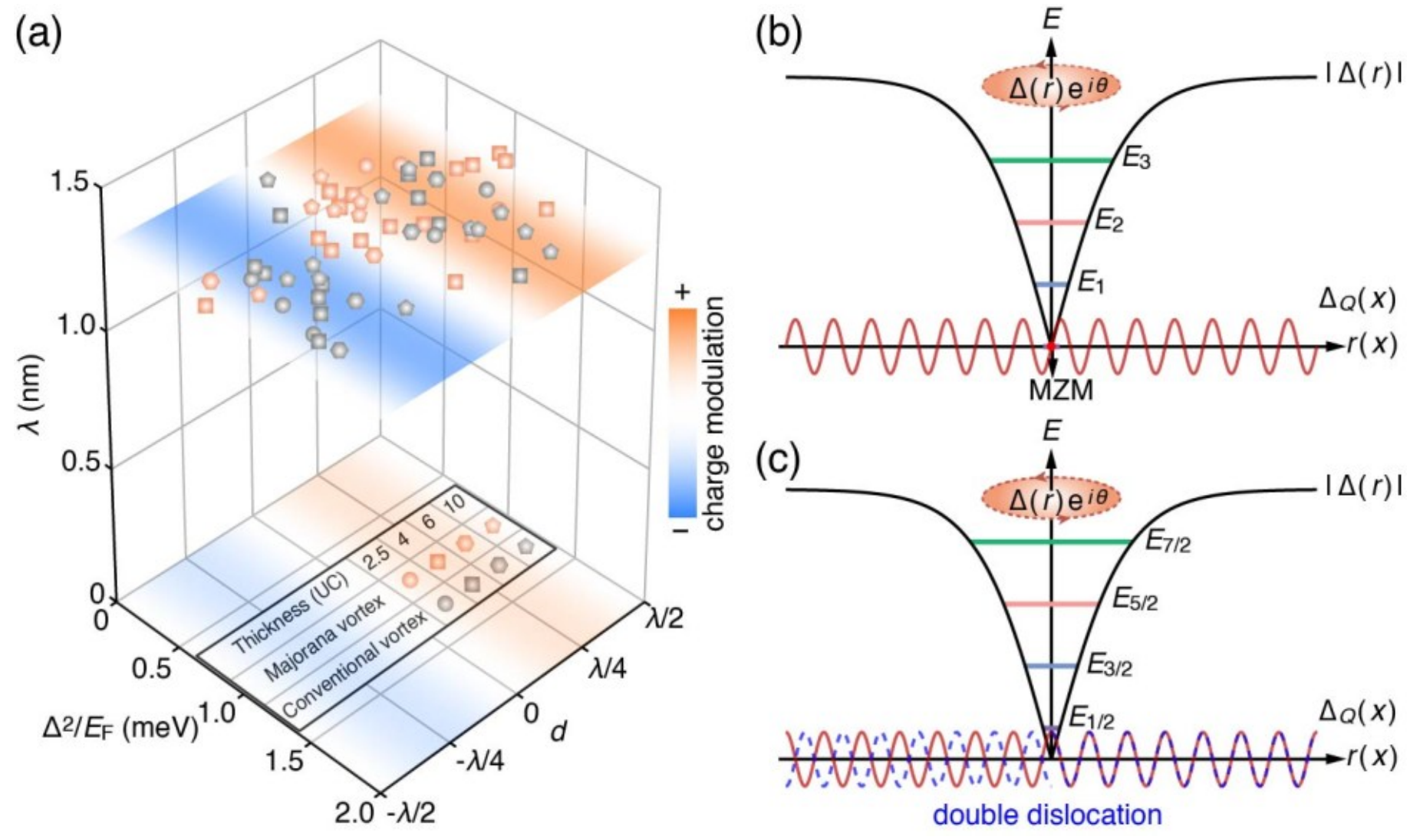

**Figure 4** (a) Three-dimensional scatter plot showing the charge-stripe wavelength $\lambda$, CdGM level spacing $\delta_E = \Delta^2/E_F$, and displacement $d$ (in units of $\lambda$) between the vortex center and the nearest charge node along the $a$ axis, measured in BFCA films of varying thickness. Each data point denotes a single vortex analyzed from high-spatial-resolution line-cut d$I$/d$V$ spectra, color-coded in orange (Majorana vortex) or gray (conventional vortex). The background schematically depicts the spatial variation of charge density, from accumulation (orange) to depletion (blue). (b,c) Schematic of Andreev bound states of Majorana and conventional vortices, localized at the nodes (b) and antinodes (c) of unidirectional $\Delta_Q(x)$ (red curves). Black curves mark the profiles of the vortexed uniform superconducting pair potential $|\Delta(r)|$, while the blue dashed curve represent the PDMs with a double dislocation at $x = 0$.

## Materials and Methods

**Sample growth.** High-quality, atomically flat $Ba(Fe_{1-x}Co_x)_2As_2$ (BFCA) films with a precisely controlled Co doping level of ~ 6.0% were epitaxially grown on 0.5wt% Nb-doped $SrTiO_3(001)$ substrates under ultra-high vacuum (UHV) conditions using state-of-the-art molecular beam epitaxy (MBE), with a base pressure better than $2.0 \times 10^{-10}$ Torr. The substrates were first degassed at 600°C for 3 hours, followed by annealing at 1250°C for 20 minutes to achieve atomically clean, flat surfaces. Prior to BFCA film growth, fluxes of high-purity Ba (99.99%), Fe (99.9999%) and Co (99.9999%) were calibrated using a quartz crystal microbalance (Inficon SQM160H). The BFCA films were subsequently grown at a substrate temperature of 500°C by co-evaporating Ba, Fe and Co at controlled stoichiometry from standard Knudsen cells under As-rich conditions (> $10^{-7}$ Torr), yielding a typical growth rate of ~ 0.17 unit cells per minute. The Co doping level and film thickness were determined from the Co/Fe flux ratio and the total deposition time. After growth, additional Ba atoms were deposited at room temperature, followed by annealing up to 600°C under reduced As pressure (~ $10^{-9}$ Torr), to form a BaAs monolayer on the FeAs surface.

**STM measurements.** All STM measurements were performed in a commercial Unisoku USM 1300 system, capable of applying a perpendicular magnetic field of up to 11 T. The base pressure in the STM chamber was maintained below $2.0 \times 10^{-10}$ Torr. High-quality BFCA films were transferred from the MBE chamber to the STM chamber through a vacuum-compatible transfer suitcase, keeping the pressure below $3.0 \times 10^{-10}$ Torr throughout the process. To further enhance the surface cleanliness for STM measurements at 0.45 K, samples were annealed in UHV at 450°C prior to insertion into the STM head. Polycrystalline PtIr tips, cleaned by *e*-beam bombardment in UHV and calibrated on MBE-grown Ag/Si(111) films, were used throughout the experiments. The STM topographic images were acquired in constant current mode, while differential conductance spectra ($dI/dV$) and maps $g(\boldsymbol{r}, V) \equiv dI/dV(\boldsymbol{r}, V)$ were measured using standard lock-in detection with a small *a.c.* modulation at a frequency of $f = 913$ Hz.

**Electrical transport measurements.** Electrical transport measurements were conducted on epitaxial BFCA films grown under identical conditions on insulating (001)-oriented $SrTiO_3$ substrates. The resistivity was measured using a standard four-terminal configuration ($I = 1\ \mu A$) in a Physical Property Measurement System (PPMS, Quantum Design). High-field $\rho_{ab}(B)$ measurements were performed at the Wuhan National High Magnetic Field Center, with the pulsed magnetic field applied perpendicular to the film surface.

**Section 1. In-plane coherence length $\xi$**

In order to determine the superconducting coherence length $\xi$, we have measured the magnetic-field-dependent resistivity $\rho_{ab}(B)$ as a function of temperature in a representative10-UC-thick sample, using pulsed fields up to 56 T applied perpendicular to the sample surface. As illustrated in Fig. S3(a), the data exhibit an apparent superconducting transition below $T_c \approx 30$ K, which is progressively suppressed with increasing field $B$. At low temperatures, the zero resistivity increases above a threshold field and eventually saturates at the normal-state value. We defined the upper critical field $B_{c2}$ as the field at which the resistivity falls to 50% of its normal-state value, and extracted $B_{c2}(T)$ at various temperatures [Fig. S3(b)]. The temperature dependence of $B_{c2}(T)$ well follows the Werthamer-Helfand-Hohenberg (WHH) model (solid line), yielding $B_{c2}(0) = 54.7$ T. Using the Ginzburg-Landau relation $\xi = (\frac{\Phi_0}{2\pi B_{c2}(0)})^2$, where $\Phi_0$ is the magnetic flux quantum, we can readily calculate the in-plane coherence length $\xi \approx 2.5$ nm, consistent with the short $\xi$ characteristic of IBSs.

**Section 2. Determination of the three parameters ($\lambda$, $\delta_E$, $d$)**

High-spatial-resolution (1.25 ~ 2.5 Å) line-cut d$I$/d$V$($r$, $E$) spectra were acquired along the $a$ and $b$ axes across the centers ($r = 0$) of 60 magnetic vortices. From these data, we first extracted the energy-independent charge-stripe wavevector $Q$ by Fourier transforming the spatially resolved conductance $g(\boldsymbol{r}, E) = dI/dV(\boldsymbol{r}, E)$ along the $a$ axis (e.g. Fig. S5(a)) at non-zero energies within $\pm\Delta_s$. To minimize measurement uncertainty, the resulting $Q$ values were averaged over various energies, from which we obtained $\lambda = 2\pi/Q$ for each vortex. Meanwhile, discrete CdGM bound states $E_\mu$ were identified from d$I$/d$V$ spectra taken along both $a$ and $b$ axes. Histograms of these $E_\mu$ values were compiled and fitted against the level index $\mu$ (integer or half-odd integer, just as in Figs. 2(d) and 2(e)), with the slope providing the CdGM level spacing $\delta_E = \Delta^2/E_F$. To determine the charge stripe-vortex registry, the vortex center was firstly determined as the coordinate at which the zero-bias conductance $g(\boldsymbol{r}, E = 0)$ reaches its maximum along the orthorhombic $a$ axis. The registry distance $d$ was then measured as the displacement between this site and the nearest nodal line of the charge-stripe modulations, normalized by the corresponding wavelength $\lambda$ to allow consistent comparison among vortices. Notably, the typical spatial resolution of 1.25 Å in our spectroscopic measurements is about one-third of the CDW node-antinode distance (~ 0.33 nm). Therefore, the uncertainty in measuring the vortex center and CDW node position may partially affect the assignment of the vortex registry, leading to the relatively broad spatial distribution of $d$ and the small fraction (< 5%) of deviating vortices shown in Fig. 4(a).

**Section 3. Nanometer-scale displacement of vortex cores and spectral switching**

To assess the robustness of the vortex-core states and their sensitivity to local environments, we carried out repeated magnetic-field applications on the same surface region of a 6-UC-thick BFCA film, marked by a native impurity at the center of the field of view [Fig. S11(a)]. After the initial field application (6 T), a vortex nucleated near the impurity exhibited a well-defined integer-quantized CdGM level sequence with a prominent zero-bias peak, characteristic of a Majorana vortex [Fig. S11(b)]. Upon complete field removal and subsequent re-application of the identical magnetic field, the same area was re-imaged. As shown in Fig. S11(c), the magnetic vortex reappeared at a slightly displaced position, having a lateral shift of approximately 1.0 nm along the *a* axis. Remarkably, the relocated vortex displayed half-odd-integer-quantized CdGM bound states without a zero-bias peak [Fig. S11(d)], indicating a transition from a Majorana to a conventional vortex. These findings demonstrate that nanoscale variations in the local environment, or in the charge stripe-vortex registry, can reversibly tune the topological character of the vortex-core states. Such reversible switching of the vortex spectra underscores the extreme sensitivity of the topological vortex-core states to the local charge environment and highlights the tunability of Majorana excitations in correlated superconductors.

**Section 4. Influence of charge-stripe order on vortex-core states**

We begin by examining the influence of the experimentally observed charge-stripe order on vortex-core states. Within a mean-field framework, the charge order can be described by a Hamiltonian [55]:

$$H_{\mathrm{CDW}}=\sum_{\boldsymbol{r}} V(1-\tanh(r/\xi))\sin[Q(x-d)]\,c_{\boldsymbol{r}}^{\dagger}c_{\boldsymbol{r}}\,,$$

where $c_{\boldsymbol{r}}^{\dagger}$ creates an electron at position $\boldsymbol{r}=(x,y,z)$ ($r=|\boldsymbol{r}|=|(x,y,z)|$ in a three-dimensional (3D) model. This term can lead to a local modulation of the chemical potential. For $d=0$, the modulation near the vortex core remains weak, whereas for $d=\pm 1/4\lambda$, the CDW introduces a finite modulation to the chemical potential, effectively shifting the Fermi energy $E_F$ of the TSS. This behavior can be understood as follows: 1) Near the vortex core, where $r$ is sufficiently small, a sine-like CDW ($d=0$) has a negligible amplitude, whereas a cosine-like modulation ($d=\pm\lambda/4$) remains finite; 2) When treated as a perturbation, the sine-type CDW gives a leading-order correction that vanishes upon spatial integration owing to the odd parity of the perturbation Hamiltonian. In contrast, the cosine-type CDW preserves even parity and thus produces a finite correction to the effective chemical potential. It is well established that in TSS-hosting IBSs, a vortex becomes topological when the $E_F$ lies below a critical value, i.e., $|E_F|<|E_{F,c}|$ where $E_{F,c}$ depends on both the bulk band structure

and the Dirac-cone parameters [50]. Thus, if a vortex is initially topological in the absence of a CDW, introducing a CDW with $d = \pm\ 1/4\lambda$ could in principle drive the system into a trivial vortex phase. However, this simple interpretation is inconsistent with our experimental observations. If a CDW-induced topological phase transition occurred when the vortex cores coincide with the charge antinodes, one would expect a strong correlation between the energy spacing $\delta_E = \Delta^2/E_F$ and the vortex types, a trend not observed experimentally [Fig. 4(a)]. This discrepancy indicates that the CDW potential scenario alone cannot account for the observed vortex dichotomy.

**Section 5. Evidence for spatially modulated superconductivity**

A closer inspection of high-spatial-resolution line-cut d$I$/d$V$ spectra and their Fourier transforms reveals that the density-wave modulations inside the vortex cores persist up to the SC gap energy [Fig. 1(f)], and that the observed leading-edge gap reduction toward the vortex centers is itself periodically modulated [Fig. 2(b)]. These features are consistent with a spatially modulated superconducting order parameter $\Delta(x)$ that coexist with the charge-stripe order in BFCA. Whereas the charge-stripe order induces a contrast reversal of $g(\boldsymbol{r}, E)$ between the occupied and empty states, the superconducting modulations manifest as particle-hole-symmetric features about $E_F$. Therefore, when they indeed coexist, symmetrizing $g(\boldsymbol{r}, E)$ with respect to $E_F$ greatly suppresses the charge-order contribution and enhances the signal of $\Delta(x)$. As shown in Fig. S12(a), symmetrization of the line-cut *dI/dV* spectra from Fig. 2(b) markedly reduces the charge modulations, leaving predominantly unmodulated vortex-core states that decay away from the vortex center. Remarkably, the superconducting gap – more precisely, the leading-edge energy – evolves periodically rather than monotonically toward the vortex center, a behavior unexpected for the uniform superconductor. The deduced leading-edge energies [Fig. S12(b)] show spatial modulation with a wavelength matching that of the charge-stripe order, pointing to its possible origin in coupling between residual charge order and re-emergent superconductivity outside the vortex core.

To further reveal the signatures of $\Delta(x)$ in BFCA, we examined zero-field d$I$/d$V$ spectra in an underdoped BFCA sample ($x \sim 0.045$). In this regime, the superconducting gaps are reduced [Fig. S13(a)], favoring the emergence of the competing charge-stripe order coupled to superconductivity, i.e., $\Delta(x)$. Strikingly, the two superconducting coherence peaks in both empty and occupied states, as well as half the distance between them (the gap magnitude $\Delta_1$), exhibits apparent spatial modulations, particularly near native impurities that further suppress the superconductivity [Fig. S13(a)]. Fourier transform of $\Delta_1(x)$ reveals a prominent peak at ~ 0.49 Å$^{-1}$ [Fig. S13(b)], matching the charge-stripe wavevector $Q$ inside vortex cores. This provides spectroscopic evidence of spatially modulated superconductivity, although a Josephson-STM measurement would be required for further identifying whether a pair density wave (PDW) exists in BFCA. The existence of $\Delta(x)$ is further supported by the distance-dependent $g(\boldsymbol{r}, E)$ modulations along the $a$ axis and its Fourier transform $g(q_a, E)$ around the gap edges [Figs. S13(c) and S13(d)]. In contrast to charge-stripe order, these $g(\boldsymbol{r}, E)$ modulations are particle-hole symmetric about $E_F$ [Figs. S13(a) and S13(c)]. The enhanced prominence of the superconducting modulations relative to the charge modulations shows that the impurity perturbs superconductivity, which would otherwise be strongly suppressed for the competing charge-stripe order to prevail in the low-energy electronic states.

**Section 6. Theoretical framework for vortex dichotomy**

The emergence of spectroscopically and topologically distinct vortex types – dictated by their nucleation near the charge nodal or antinodal lines inside the vortices – is difficult to reconcile with a superconductivity described solely by a uniform component $\Delta_0$. Such a scenario would also be unnatural: in the presence of a uniform SC order parameter $\Delta_0$, a CDW order $\rho_Q^{\mathrm{CDW}}(x) = \rho_0\sin(Qx + \phi_C)$ with $Q = 2\pi/\lambda$ necessarily induces a spatially modulated superconducting component $\Delta_Q(x) \propto \Delta_0\rho_Q^{\mathrm{CDW}}(x) = \Delta_P\sin(Qx + \phi_P)$ with $\Delta_P \sim \Delta_0\rho_0$ [Figs. 4(b) and 4(c)] [52,53], where $x$ is the coordinate along the $a$ direction. The CDW-induced superconducting modulations have been observed in the SC state of the prototypical CDW-hosting $NbSe_2$ [54]. In general, a relative phase difference ($\phi_C \neq \phi_P$) may exist between the CDW and $\Delta_Q(x)$. However, given the pronounced CDW modulations observed inside the vortex cores, we reasonably assume an in-phase relation between them ($\phi_C = \phi_P$), such that the nodal lines of $\Delta_Q(x)$ lie at $x = \pm\frac{\lambda}{2}n$ ($n$ = 0, 1, 2, …) along the Fe-Fe direction for the Majorana vortices [Fig. 4(b)], and at $x = \pm\frac{\lambda}{2}n + \frac{\lambda}{4}$ for conventional vortices [Fig. 4(c)]. Considering the presence of $\Delta_Q(x)$, the SC order parameter around an isolated vortex can be expressed as a superposition of the vortexed uniform component and a unidirectional term, i.e., $\Delta_{\mathrm{vortex}}(r) = |\Delta(r)|e^{i\nu\theta} + \Delta_P\sin[Q(x - d) + \varphi(x)]$, where $|\Delta(r)| \approx \Delta_0\tanh(r/\xi)$ is the amplitude of the vortexed SC order parameter centered at $r$ = 0, around which the phase winds by $2\pi$ times the vorticity $\nu$ = 1, and $\varphi(x)$ represents possible dislocations in $\Delta_Q(x)$.

Because the amplitude $|\Delta(r)|$ is suppressed inside the vortex cores, the $\Delta_Q(x)$ component becomes prominent and can substantially modify the spectrum of the vortex-core states.

The coexistence of vortices and PDWs, described by a similar SC order parameter, has been theoretically explored in connection with PDW pinning in the vortex halos of high-$T_c$ cuprates [58,59]. In our case, the physics is further enriched by the presence of SC TSS, which profoundly modify the vortex-core spectrum. When a vortex nucleates at the $\Delta_Q(x)$ nodes ($d$ = 0, Fig. 4(b)), the modulated superconducting component take the form $\Delta_Q(x) \propto \Delta_P \sin[Qx + \varphi(x)]$. Here, it suffices to set the dislocation phase factor $\varphi(x) = 0$, corresponding to an undistorted $\Delta_Q(x)$ that is mirror-odd under reflection across the nodal line at $x = 0$. This symmetry matches the mirror-odd character of the vortex order parameter $|\Delta(r)|e^{i\theta}$, which arises from the sign change of the phase-winding factor [Fig. 4(b)]. For such vortices, minimal alteration of the vortex-core states is expected. Indeed, a simple model calculation of the CdGM spectrum for a vortex associated with the SC TSS and centered at the $\Delta_Q(x)$ nodes [Fig. S14(a,b), Sec. 7] confirms integer quantization of the vortex-core levels with robust MZMs.

This situation changes markedly when the vortex nucleates at the $\Delta_Q(x)$ antinodes ($d = \pm\frac{\lambda}{4}$, Fig. 4(c)), for which $\Delta_Q(x) = \Delta_P \cos[Qx + \varphi(x)]$. With $\varphi(x) = 0$, the undistorted $\Delta_Q(x)$ is mirror-even with respect to the $x = 0$ plane [see red curve in Fig. 4(c)], rendering it incompatible with the mirror-odd vortex order parameter $|\Delta(r)|e^{i\theta}$. This incompatibility necessitates the formation of dislocations, as discussed theoretically [58-61]. For instance, a double dislocation can arise at $x = 0$, corresponding to $\varphi(x) = 0$ for $x > 0$ and $\varphi(x) = \pi$ for $x < 0$ [see blue curve in Fig. 4(c)], thereby restoring mirror-odd symmetry of the $\Delta_Q(x)$ to match the vortex order parameter. Our model calculations show that the phase discontinuity associated with such a double dislocation strongly perturbs vortex-core states of the SC TSS [Fig. S14(c), Sct.7], destabilizing the integer quantization sequence and removing the MZM. Notably, the spatial modulation of the superconducting order parameter in our calculations is introduced phenomenologically to examine its symmetry matching with the vortex states, rather than being obtained from a fully self-consistent microscopic calculation. Therefore, the present model should be regarded as a qualitative interpretation of the observed vortex dichotomy. A fully self-consistent description of the coupled $\Delta_Q(x)$, CDW, and vortex structure remains an important subject for future theoretical investigation.

**Section 7. Model calculations**

To phenomenologically describe the CdGM states inside a vortex in the presence of $\Delta_Q(x)$, we adopt a 2D minimal model, where the vortex of SC TSS hosts the zero-energy MZM in the absence of density-wave modulations and the CDW contribution is too weak to drive a topological phase transition. The effective Bogoliubov-de Gennes (BdG) Hamiltonian then reads [56]

$$H = \frac{1}{2}\int d^2r\Psi^\dagger \begin{pmatrix} \hat{h} & \hat{\Delta} \\ \hat{\Delta}^* & -\hat{h} \end{pmatrix} \Psi,$$

where $\hat{h} = v_F \boldsymbol{\sigma} \cdot \boldsymbol{p} - E_F$, with $\boldsymbol{\sigma}$ the Pauli matrices in the spin basis and $v_F$ the Fermi velocity associated with the spin-orbital coupling. The superconducting gap operator is

$$\hat{\Delta} = (\Delta_0 \tanh(r/\xi) \exp(i\theta) + \Delta_Q(x))\sigma_0,$$

With $(r, \theta)$ the polar coordinates. When $\Delta_Q(x) = 0$, the vortex exhibits topological character and supports a MZM inside its core, as shown in Fig. S14(a). The resulting discrete energy levels follow the quantization $E_\mu = \mu\Delta^2/E_F$ ($\mu$ is the angular momentum quantum number) [50]. Here we performed the calculation on a 90 × 90 lattices with a coherence length $\xi = 10$, ensuring that the edge-state boundary is well separated from the vortex-core-localized MZM.

Next, we include the spatially modulated superconducting term defined as $\Delta_Q(x) = \Delta_P \sin[Q(x - d) + \varphi(x)]$, where $\varphi(x)$ incorporates the effect of dislocations [61,62]. The vortexed uniform superconducting order is suppressed within the vortex core, so that the spatially modulated superconducting term becomes energetically relevant in the Ginzburg-Landau free energy [58]:

$$F_q \sim \int d^2r[\Delta^*(r)\Delta_Q(x) + \text{c.c.}],$$

where $\Delta(r)$ is odd under spatial inversion, i.e., $\Delta(-r) = -\Delta(r)$. To minimize the free energy $F_q$, the finite-momentum contribution must remain nonvanishing. For $d = 0$, $\Delta_Q(x)$ is also odd under inversion with $x = 0$, yields a nonzero overlap integral. However, for $d = \pm 1/4\lambda$, this symmetry is lost, necessitating introduction of dislocation. To restore the odd parity under space reversal when $d = \pm 1/4\lambda$, a double dislocation must occur near $x = 0$, implemented as $\varphi(x) = \pi\Theta(-x)$, with $\Theta(x)$ the Heaviside step function. Under these considerations, we compute the energy spectra of a vortex coexisting with $\Delta_Q(x)$ for both $d = 0$ and $d = \pm 1/4\lambda$, shown in Fig. S14(b,c). The results reveal that a $\Delta_Q(x)$ with $d = 0$ preserves the MZM, while a double dislocation at the $x = 0$ plane [Fig. 4(c)] for $d = 1/4\lambda$ effectively suppresses the zero-energy mode. This reveals that the superconducting phase dislocations can influence the vortex-core spectra and destabilize the MZM, consistent with our experimental observations. The calculations were performed with the software package kwant [57]. To fully capture the observed properties of the vortex core states, it is necessary to include both the charge-stripe modulations in the particle-hole sector and the spatially modulated superconducting order parameter in the particle-particle sector, which will be studied in the future.

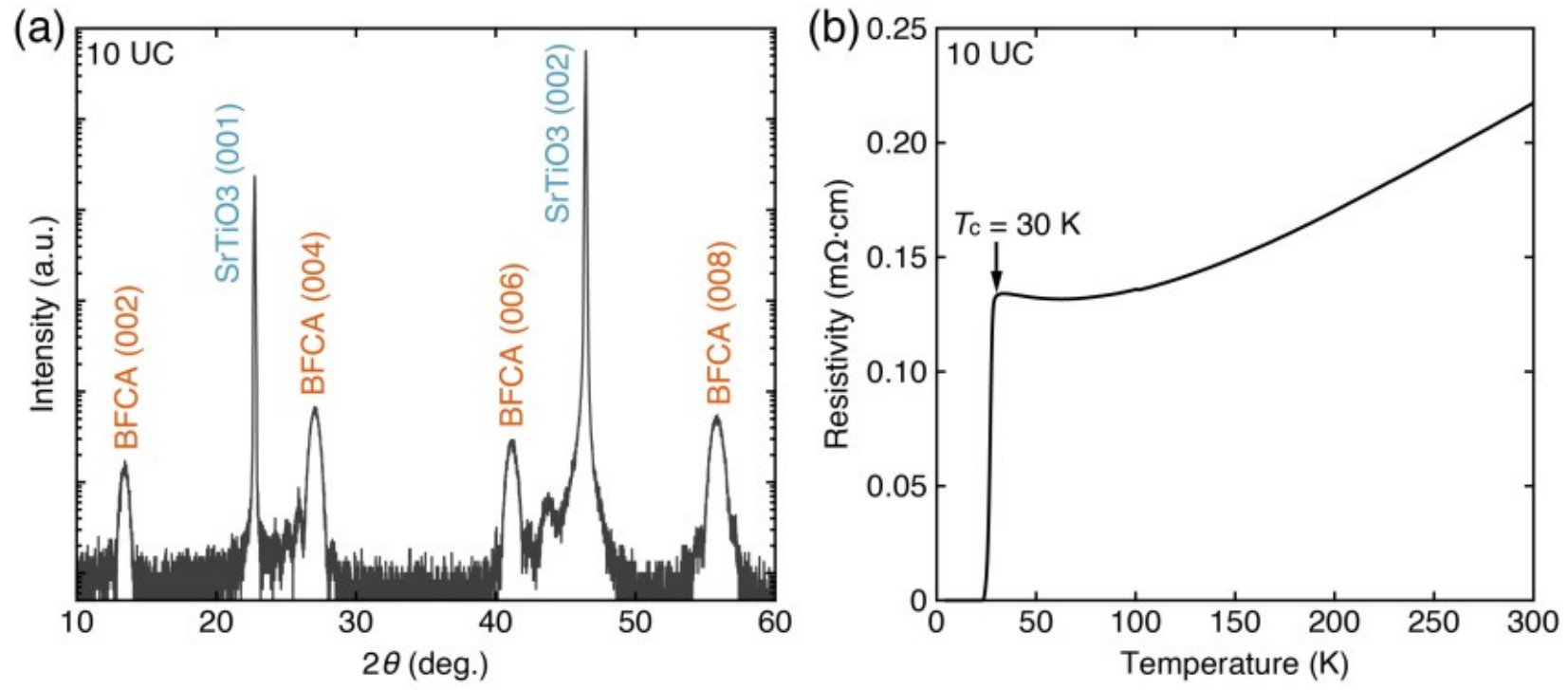

**Fig. S1.** (a) X-ray diffraction (XRD) spectrum measured in a representative 10-UC-thick compressively strained BFCA film ($c \approx 13.2$ Å) on $SrTiO_3(001)$ substrate, using monochromatic copper $K_{\alpha 1}$ radiation with a wavelength of 1.5406 Å. The pronounced BFCA(00$n$) reflections and Laue oscillations around them indicate superior crystallinity of the films. (b) Temperature-dependent in-plane electrical resistivity $\rho_{ab}$ for an epitaxial BFCA film near optimal Co (~ 6.0%) doping, exhibiting a superconductivity transition with $T_c \approx 30$ K.

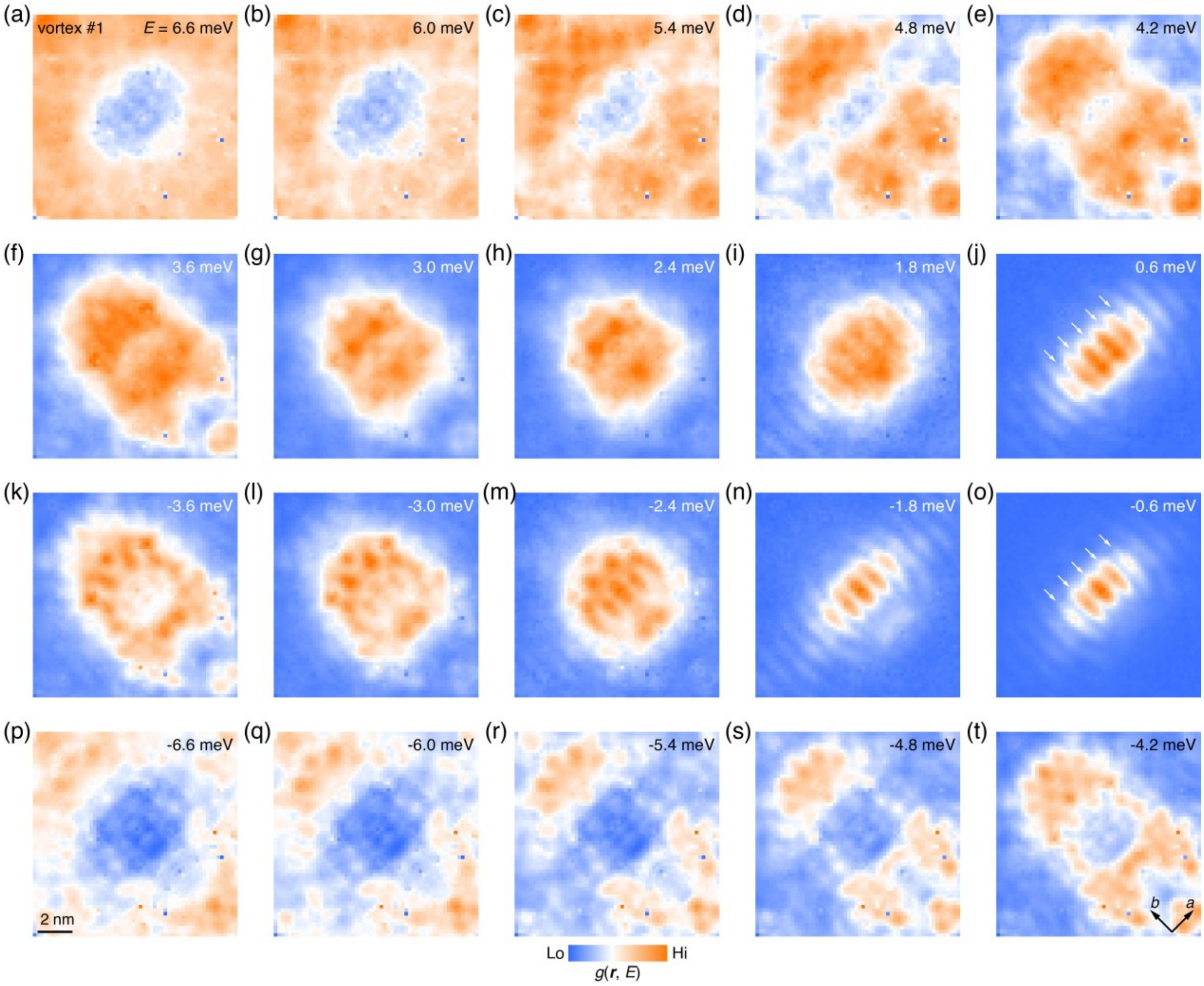

**Fig. S2.** (a-t) Energy-resolved conductance maps $g(\mathbf{r}, E)$ measured in the same field of view of Fig. 1(c). The *a*-axis charge modulations are pronounced within the energy window of ± 3.0 meV (i.e., ~ $\pm\Delta_s$).

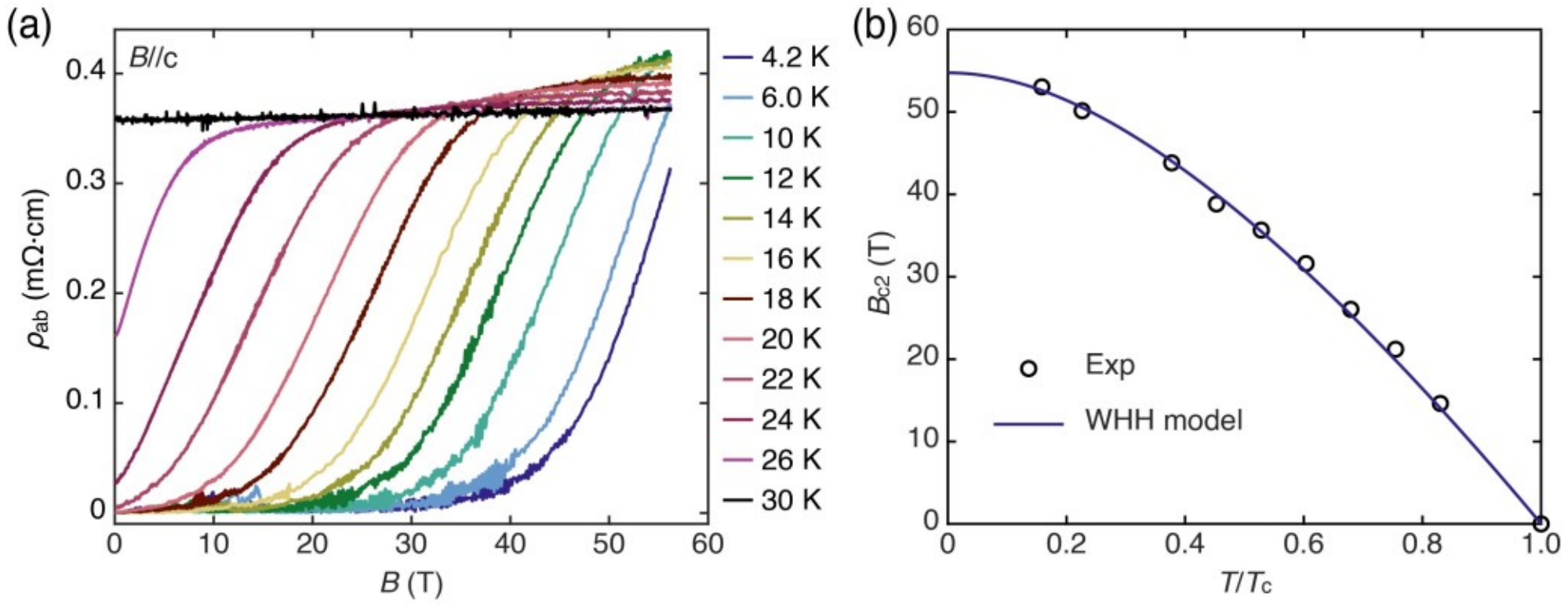

**Fig. S3.** (a) Magnetic-field-dependent resistivity $\rho_{ab}(B)$ of a 10-UC-thick BFCA film measured under pulsed fields up to 56 T at various temperatures, showing progressive suppression of superconductivity with increasing field. (b) Upper critical field $B_{c2}$ plotted as a function of normalized temperature $T/T_c$. The solid line represents a fit of the data (circles) to the Werthamer-Helfand-Hohenberg model.

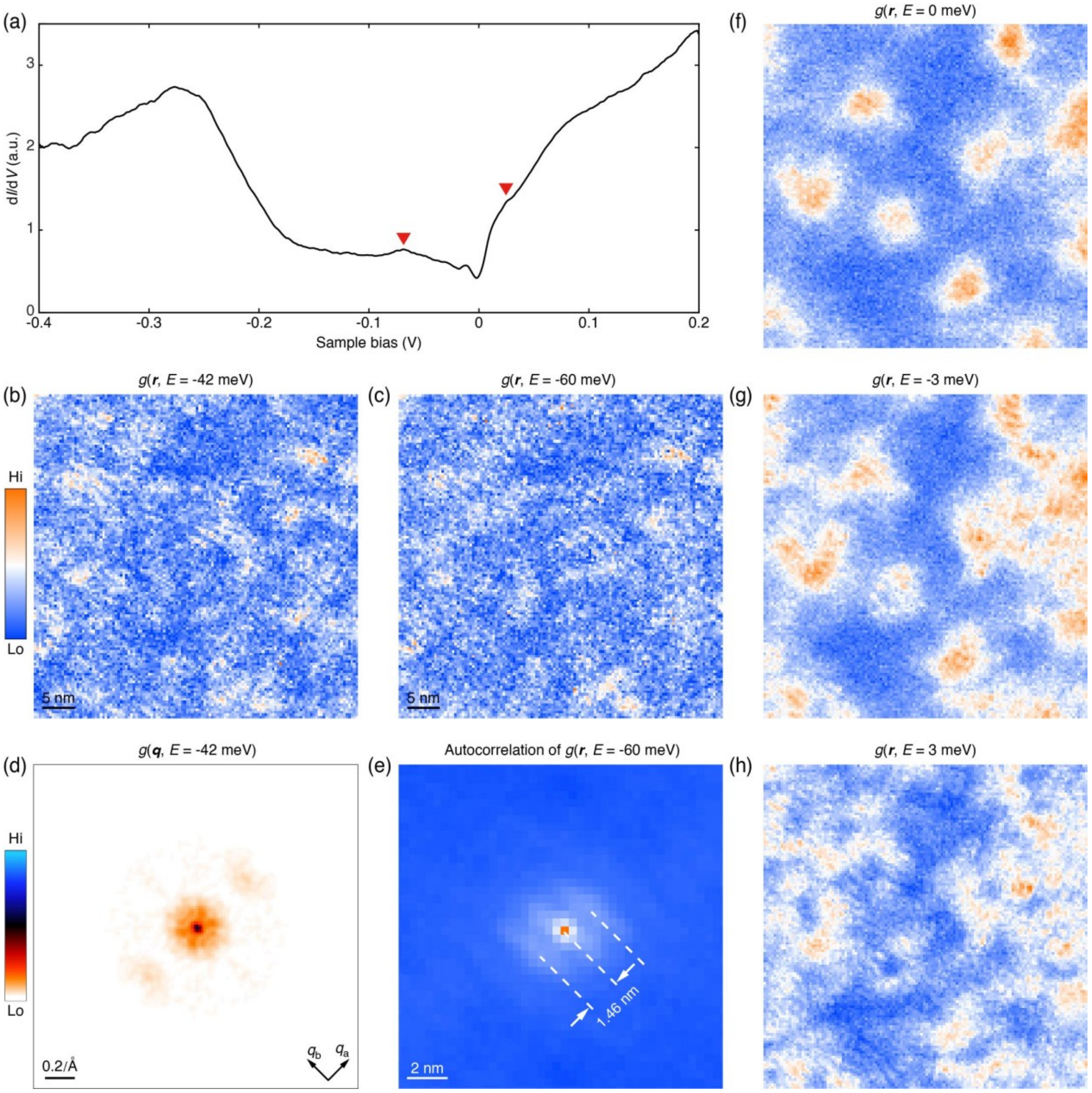

**Fig. S4.** (a) Representative d$I$/d$V$ spectrum in the wide energy range, showing gap-edge-like features beyond the superconducting gap (marked by red triangles). (b,c) Energy-resolved $g(\boldsymbol{r}, E)$ maps measured over a 50 nm × 50 nm field of view in a 10-UC-thick film, revealing weak but discernible stripe-like electronic modulations at energies outside the superconducting gap. Setpoint: $V$ = 60 mV, $I$ = 0.15 nA. (d) Fourier-transform amplitude $g(\boldsymbol{q}, E)$ obtained from the $g(\boldsymbol{r}, E)$ in (b), showing a unidirectional modulation feature at $Q \sim 0.45 \pm 0.03$ Å$^{-1}$ along the $q_a$ direction. (e) Autocorrelation analysis of the $g(\boldsymbol{r}, E)$ map in (c), yielding a modulation period comparable to that observed for the stripe modulations within vortex cores. (f-h) Energy-resolved $g(\boldsymbol{r}, E)$ maps acquired in the same field of view at 8 T and 4.2 K, revealing similar spatially modulated electronic states within the vortex cores.

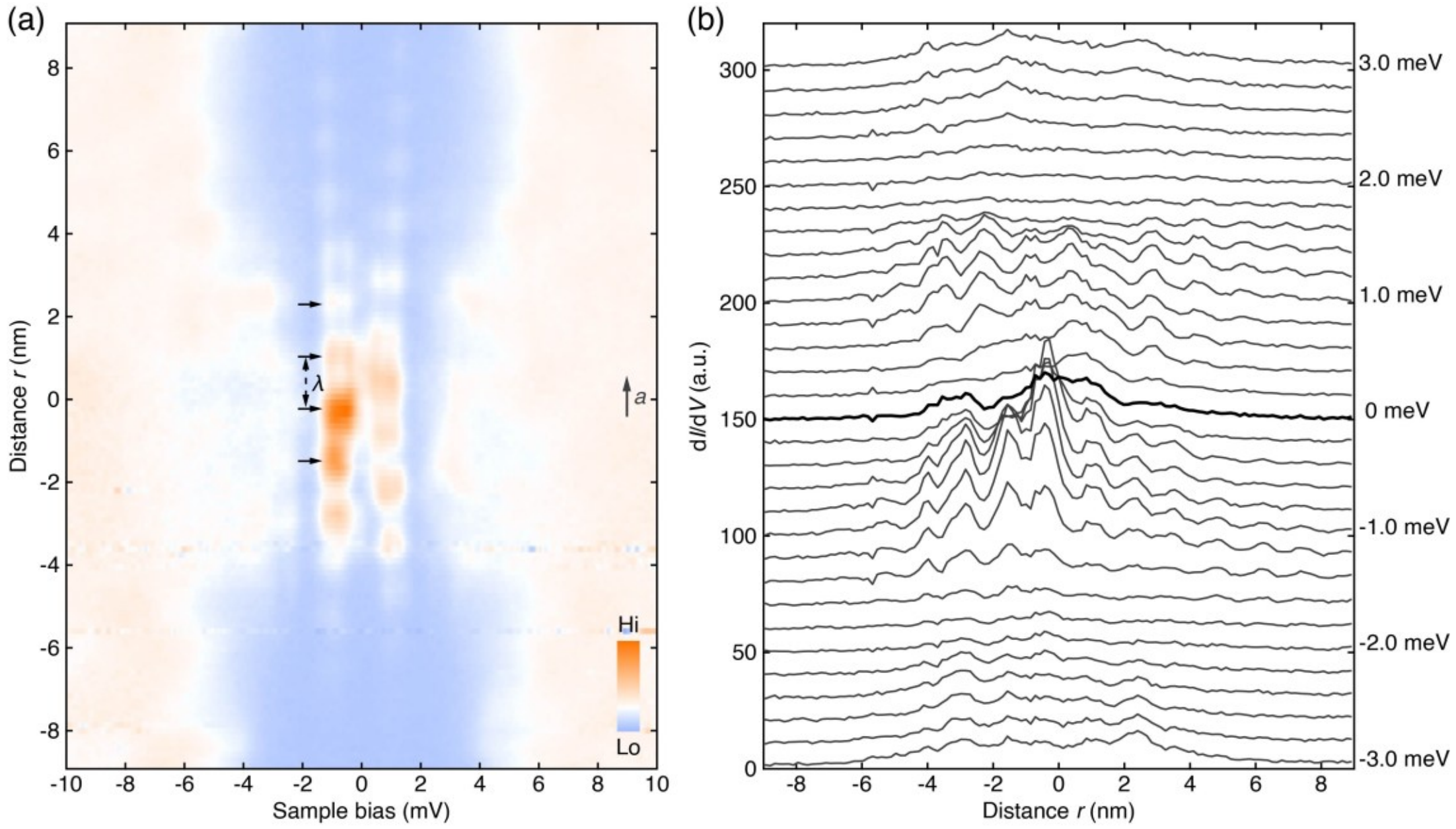

**Fig. S5.** (a) Intensity plot of line-cut d$I$/d$V$ spectra across the center ($r$ = 0) of Majorana vortex #1 along the orthorhombic $a$ axis. Black arrows denote the charge-stripe modulations. Setpoint: $V$ = 10 mV, $I$ = 0.12 nA. (b) Spatially resolved $g(\boldsymbol{r}, E)$ profiles acquired across the center of vortex #1 along the $a$ axis, with energies ranging from -3.0 meV (bottom) to 3.0 meV (top). The thick curve represents $g(\boldsymbol{r}, E)$ at $E$ = 0. These profiles exhibit clear spatial modulations, with the contrast reversing between $E$ < 0 and $E$ > 0, while maintaining the same phase below (or above) $E_F$.

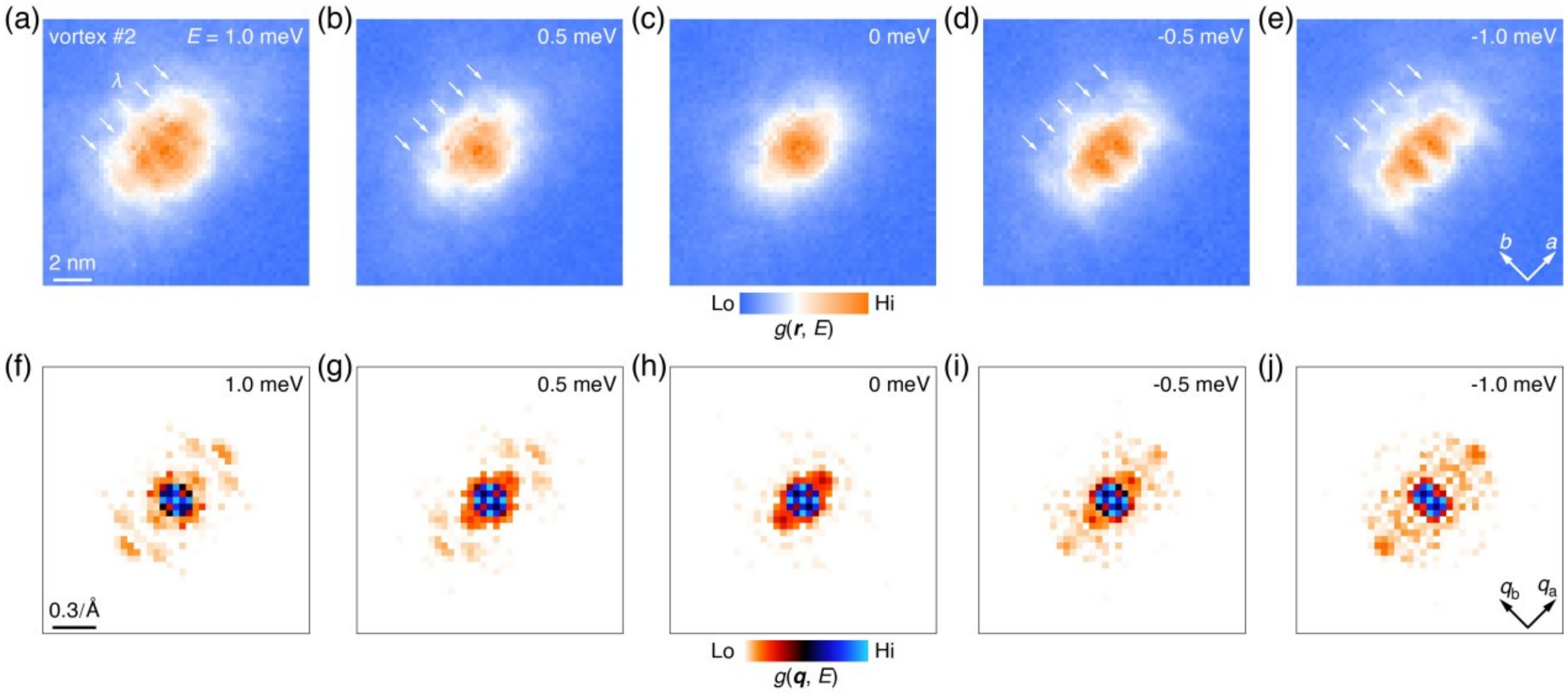

**Fig. S6.** (a-e) Energy-resolved $g(\boldsymbol{r}, E)$ maps measured over a 13.5 nm × 13.5 nm field of view at $B$ = 6 T, revealing unidirectional charge stripes (white arrows) along the $a$ axis within the halo of vortex #2. Note that the charge-induced spatial modulations of $g(\boldsymbol{r}, E)$ are in-phase either above or below $E_F$ in contrast, but undergo a π-phase shift upon reversal of the sample-bias polarity. Setpoint: $V$ = 10 mV, $I$ = 0.1 nA. (f-j)

Amplitude $g(\boldsymbol{q}, E)$ maps derived from Fourier transforms of $g(\boldsymbol{r}, E)$ in (a-e), exhibiting scattering intensity predominantly at non-zero energies at $Q = (\sim 0.45$ Å$^{-1}$, 0).

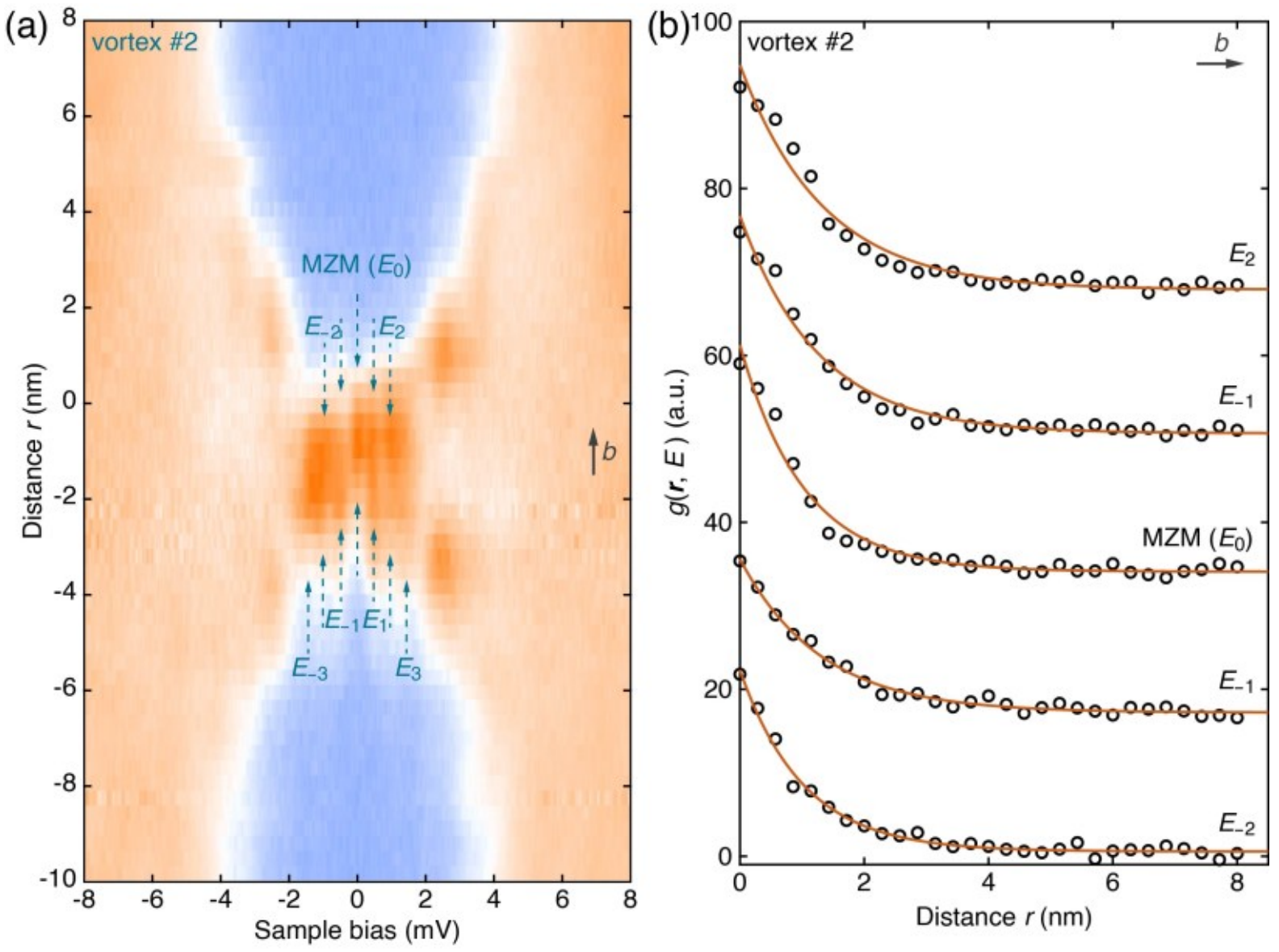

**Fig. S7.** (a) Intensity plot of line-cut *dI/dV* spectra across Majorana vortex #2 along the orthorhombic *b* axis, displaying no periodic spatial modulation. Green arrows indicate the discrete CdGM bound states. Setpoint: $V$ = 10 mV, $I$ = 0.15 nA. (b) Spatially resolved $g(\boldsymbol{r}, E)$ profiles extracted from (A) at the discrete $E_\mu$ energies, as indicated. In contrast to the oscillatory behavior along the *a* axis, all $g(\boldsymbol{r}, E)$ profiles exhibit a purely monotonic exponential decay away from the vortex center, as shown by the solid lines.

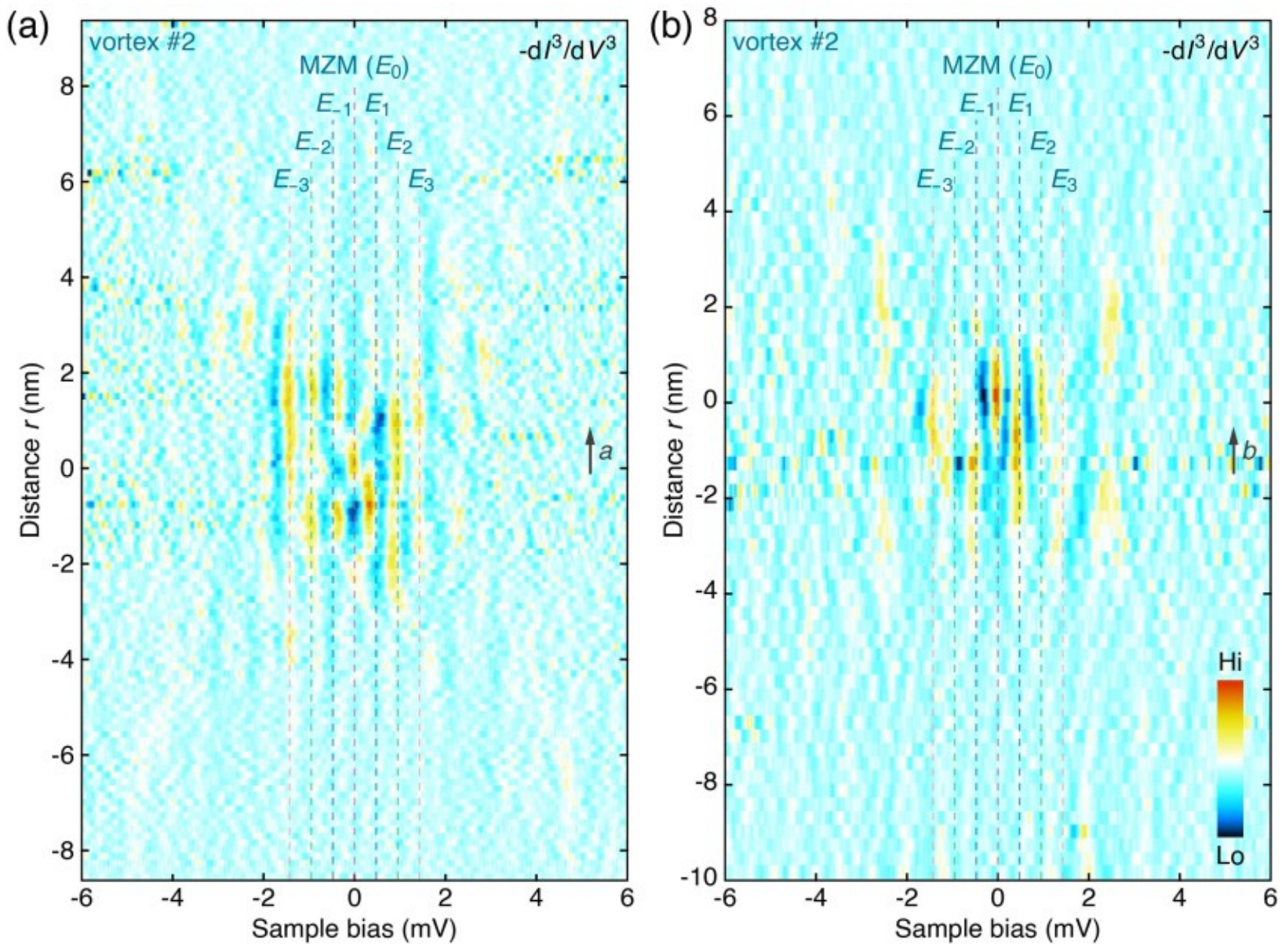

**Fig. S8.** (a,b) Negative second derivative of the line-cut d$I$/d$V$ intensities (i.e. -d$I^3$/d$V^3$) across the center of vortex #2 along the orthorhombic *a* and *b* axes, respectively. The discrete CdGM bound states in the original

d$I$/d$V$ spectra are strongly enhanced as sharp peaks in -d$I^3$/d$V^3$. Notably, neither the zero-energy MZM nor the finite-energy CdGM bound states exhibit spatial splitting away from the vortex center (vertical dashed lines). This derivative analysis gives a robust and independent method for extracting the energy positions $E_\mu$ of discrete CdGM bound states, yielding values that match with those obtained from multi-Gaussian fits.

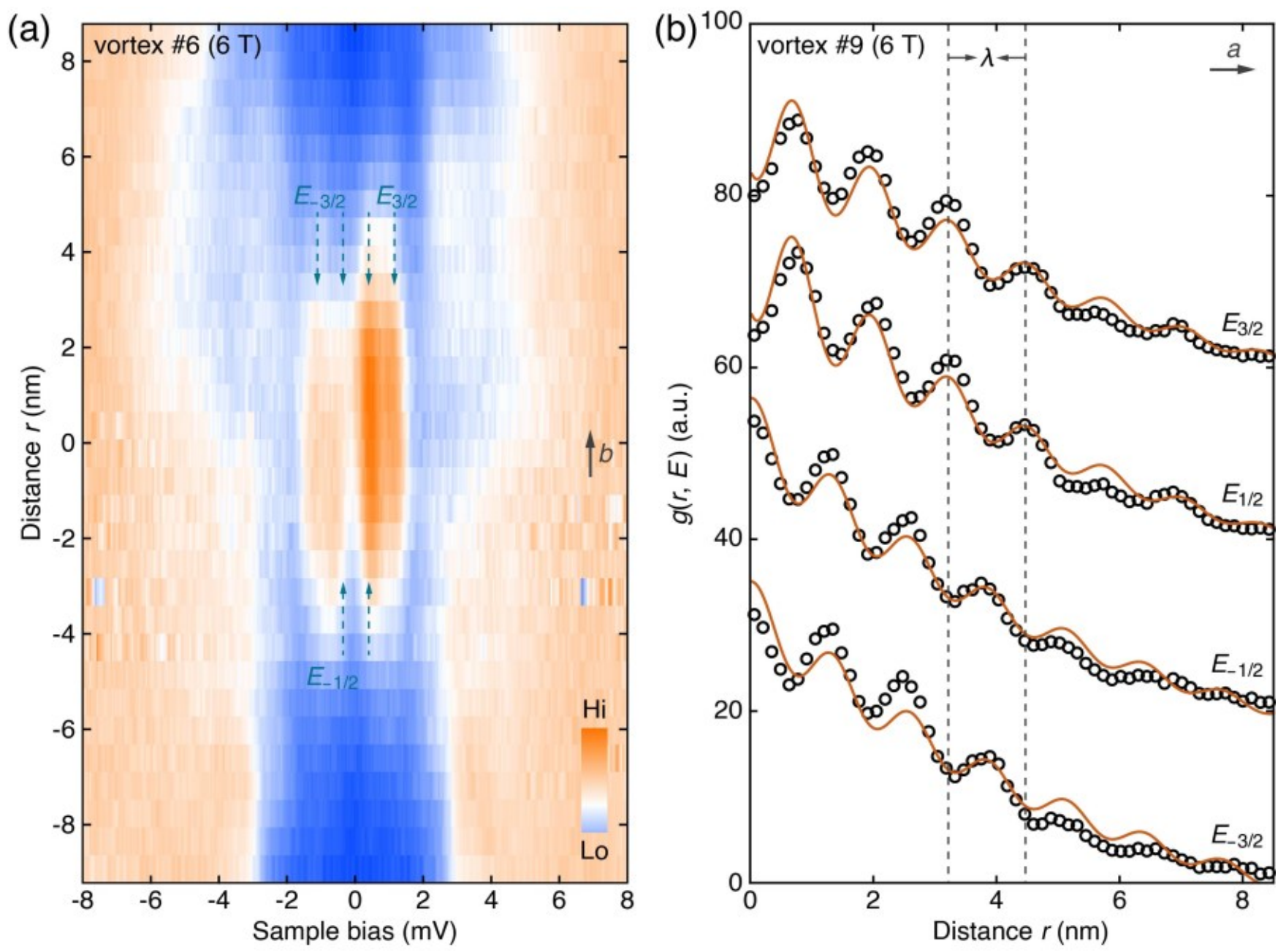

**Fig. S9.** (a) Intensity plot of line-cut d$I$/d$V$ spectra across the conventional vortex #6 in 10-UC-thick films, measured along the $b$ axis. Green arrows indicate the discrete CdGM bound states at half-integer energy levels $E_\mu$ ($\mu$ = ±1/2, ±3/2). Setpoint: $V$ = 8 mV, $I$ = 0.25 nA. (b) Spatially resolved $g(\boldsymbol{r}, E)$ profiles, acquired along the $a$ axis at the discrete $E_\mu$ energies, as indicated. The CdGM bound states exhibit spatial modulations in intensity with a wavelength $\lambda \sim 1.27$ nm. Solid curves represent the best fits of $g(\boldsymbol{r}, E)$ to oscillatory decay functions.

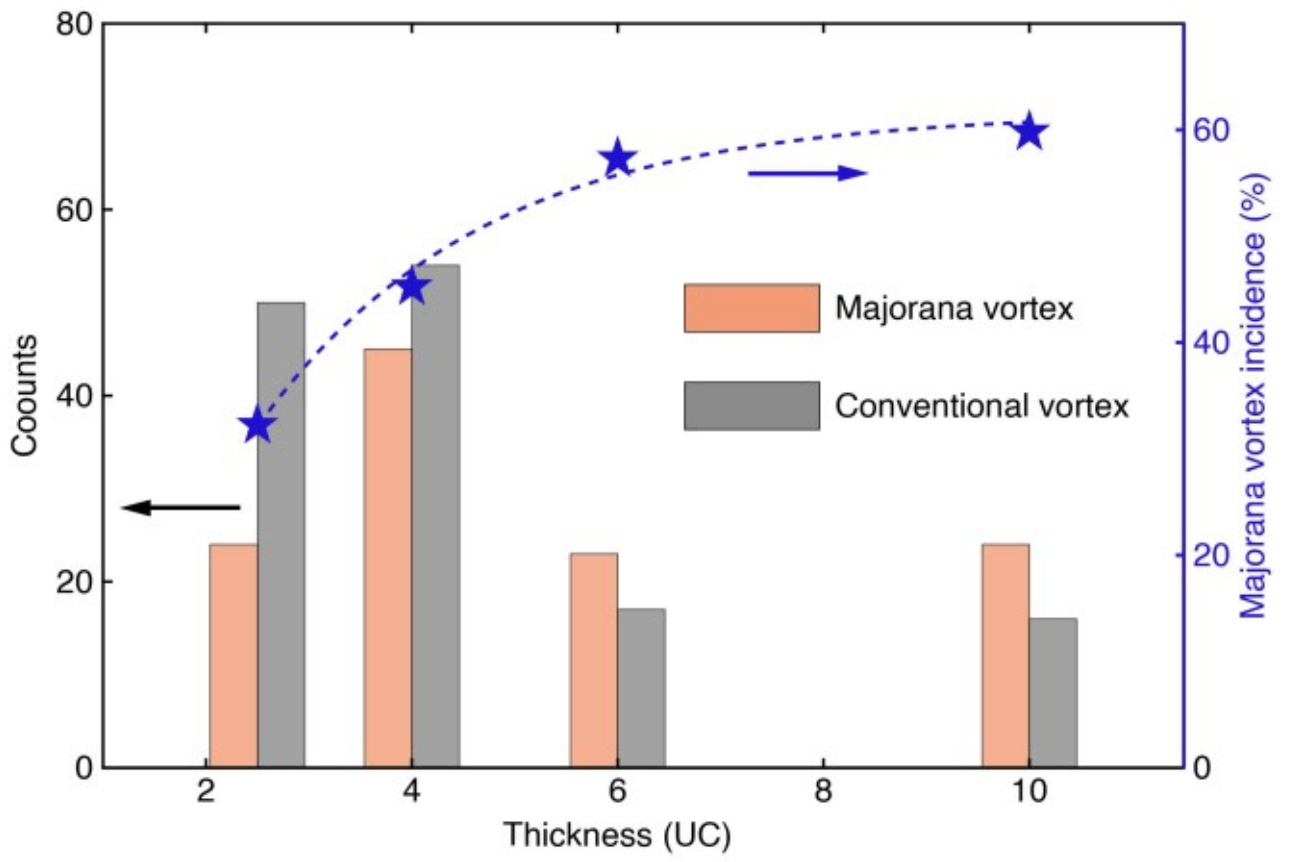

**Fig. S10.** Histogram (left axis) summarizing the distribution of Majorana (orange) and conventional (gray) vortices studied, based on the CdGM level quantization (integer *versus* half-integer) and the presence of MZMs. The right axis displays the thickness-dependent incidence of Majorana vortices in BFCA films of varying thickness. In 10 UC films, approximately 60% of vortices contain MZMs, but this fraction drops to ~ 32% in 2.5 UC samples. The reduced Majorana vortices in thinner BFCA highlights the sensitivity of charge stripe-vortex registry to dimensionality effects.

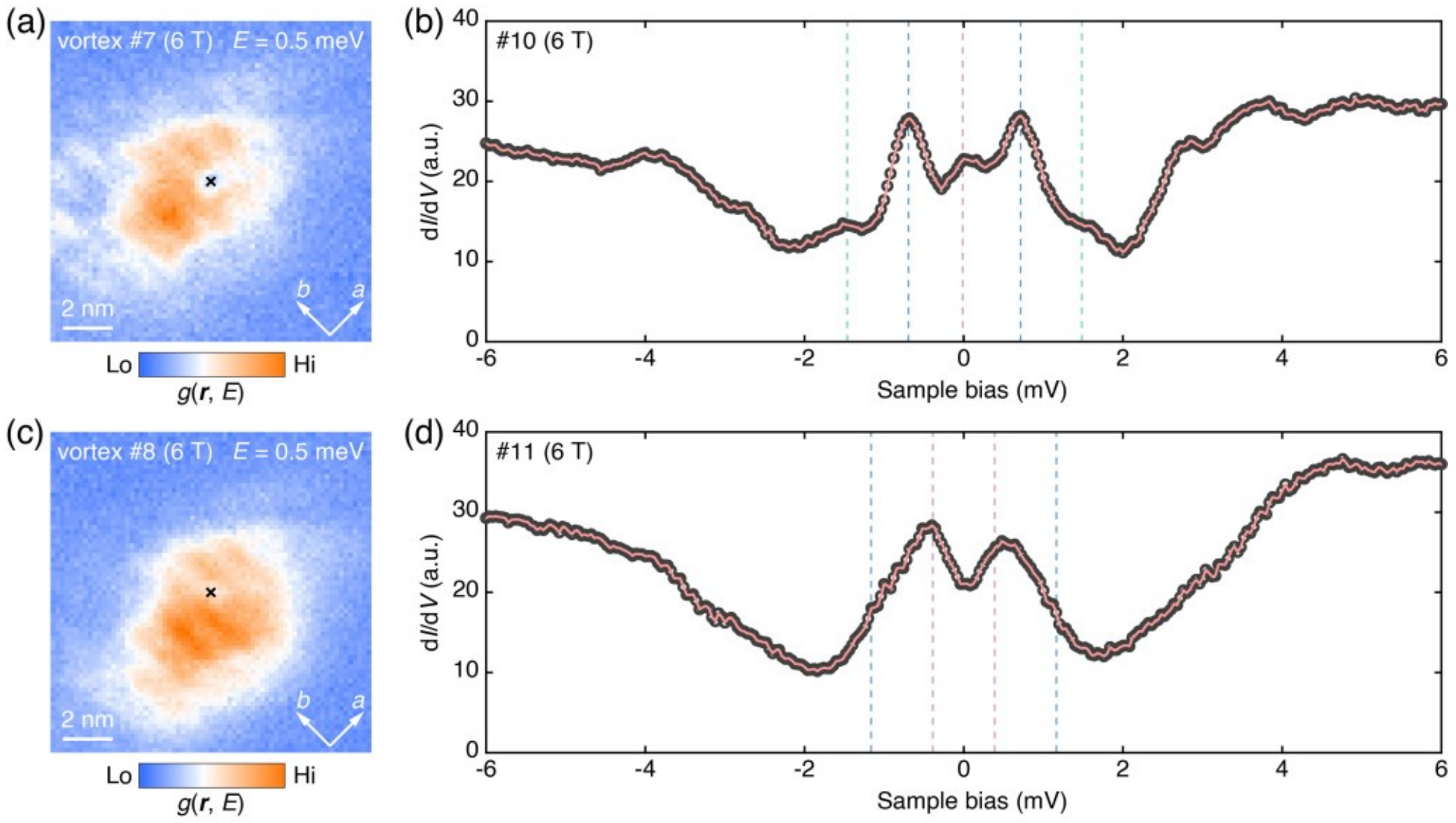

**Fig. S11.** (a,b) Conductance map $g(\boldsymbol{r}, E = 0.5$ meV) and representative d$I$/d$V$ spectrum acquired around Majorana vortex # 7 in 6-UC-thick films, showing integer-quantized CdGM levels accompanied by a zero-bias conductance peak. (c) $g(\boldsymbol{r}, E = 0.5$ meV) map measured in the same field of view (12.9 nm × 12.9 nm) as (a) after removing and reapplying an identical magnetic field of 6 T. The vortex cores reappeared at a slightly displaced position, as referenced to the cross-marked native impurity. (d) Representative d$I$/d$V$

spectrum acquired inside the relocated magnetic vortex #8 in (c), exhibiting half-odd-integer quantization without a zero-bias peak. Setpoint: $V = 8$ mV, $I = 0.2$ nA.

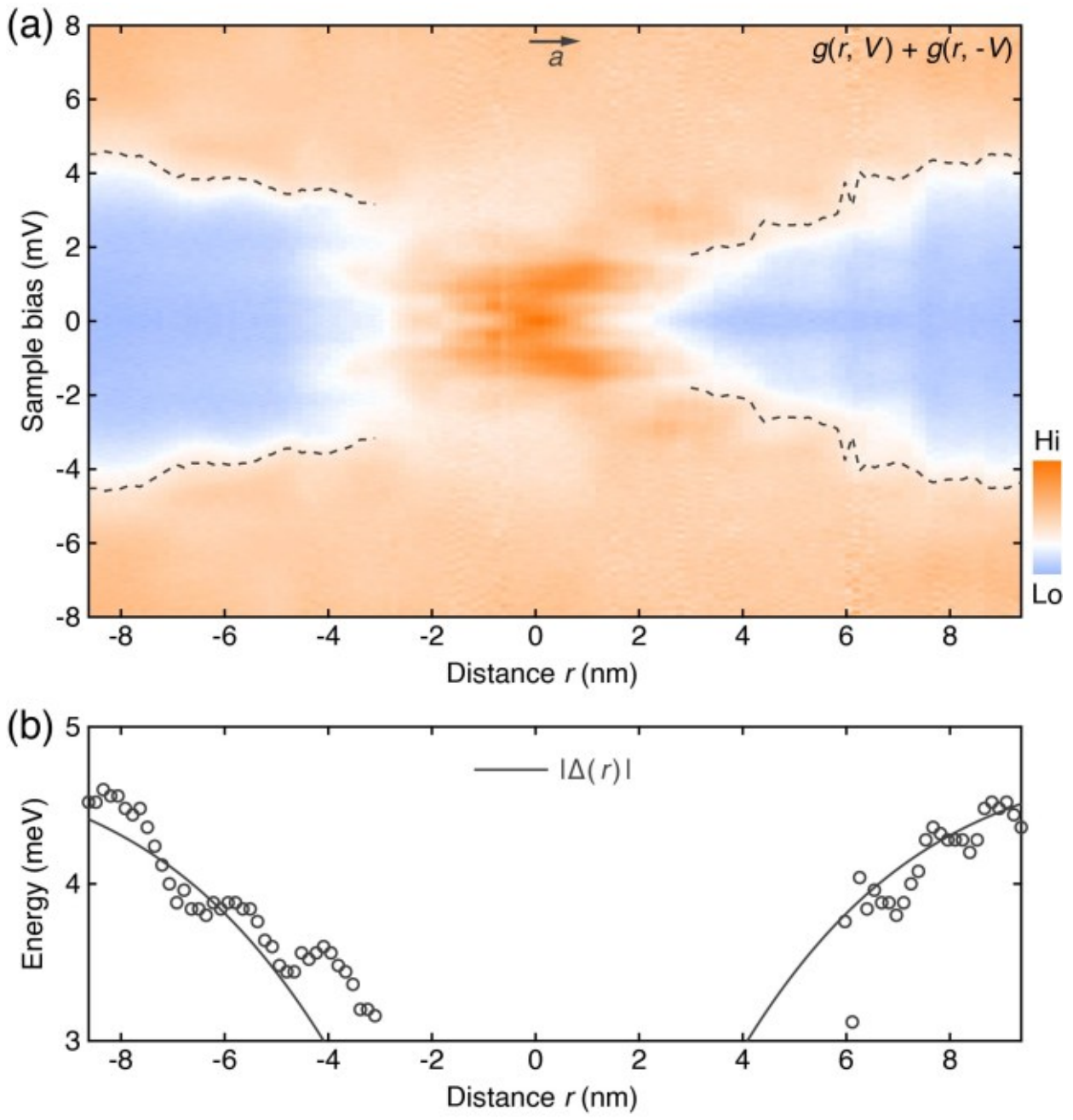

**Fig. S12.** (a) Symmetrization of the line-cut d$I$/d$V$ spectra in Fig. 2(b) with respect to $E_F$, showing modulated leading-gap edges outside the vortex cores around vortex #2 (marked by dashed lines). (b) Leading-edge energy (circles) as a function of distance from the vortex center. In contrast to the monotonical variation of the vortexed uniform superconducting pair potential $|\Delta(r)|$ (black curves), the leading-edge energy exhibits a periodic modulation with a wavelength matching the charge-stripe wavelength $\lambda \sim 1.30$ nm.

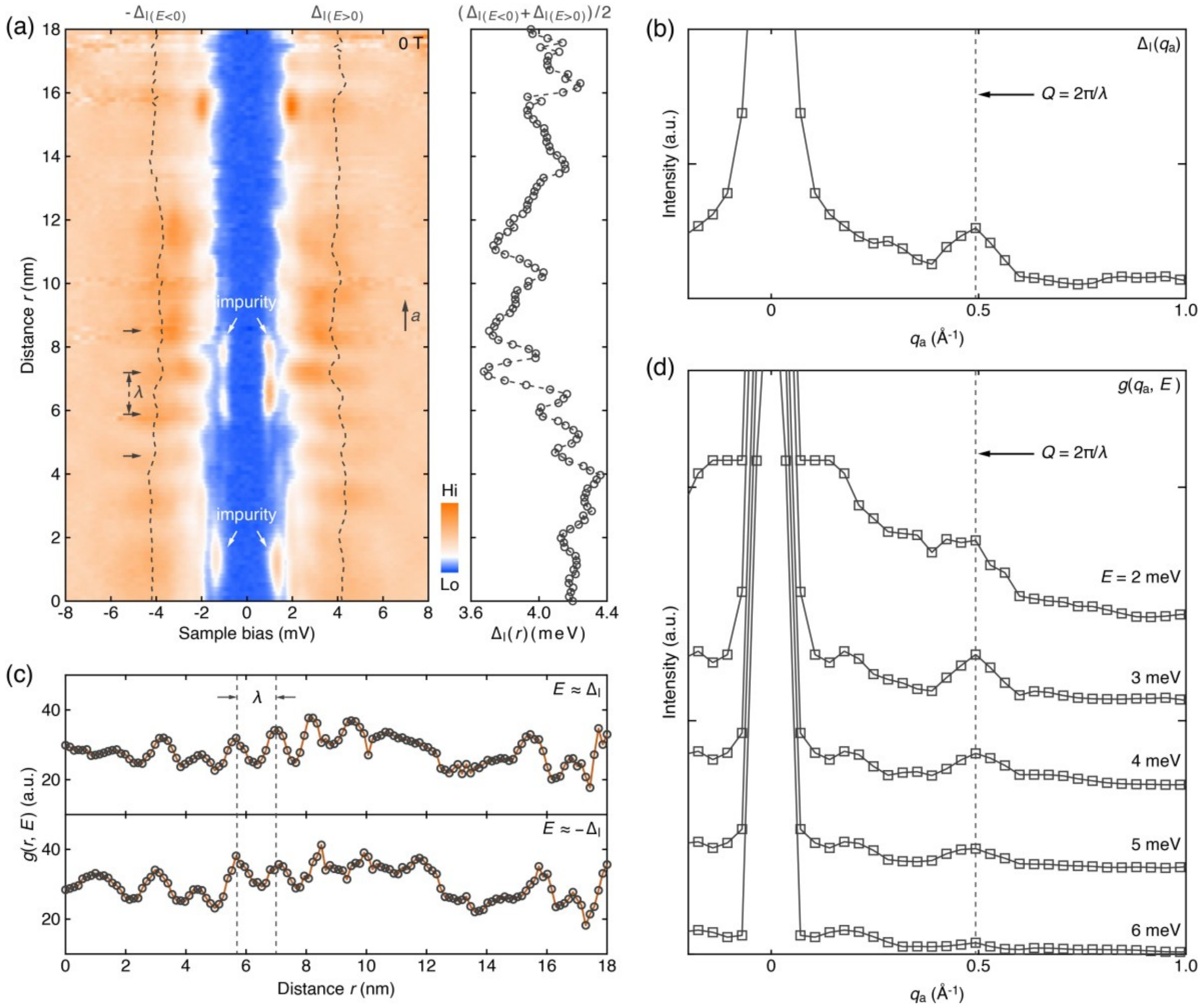

**Fig. S13.** (a) Intensity plot of zero-field line-cut d$I$/d$V$ spectra taken along the $a$ axis in an underdoped BFCA film ($x \approx 0.045$), showing spatial modulation of the superconducting gap magnitude $\Delta_l$. (b) Fourier transform of $\Delta_l(r)$, showing a sharp peak at $Q \approx 0.49$ Å$^{-1}$. (c) Distance-dependent $g(\boldsymbol{r}, E)$ modulations around the two coherence peaks ($\pm\Delta_l$), which are particle-hole symmetric about $E_F$. (d) Fourier transforms $g(q_a, E)$ at various energies around the gap edges.

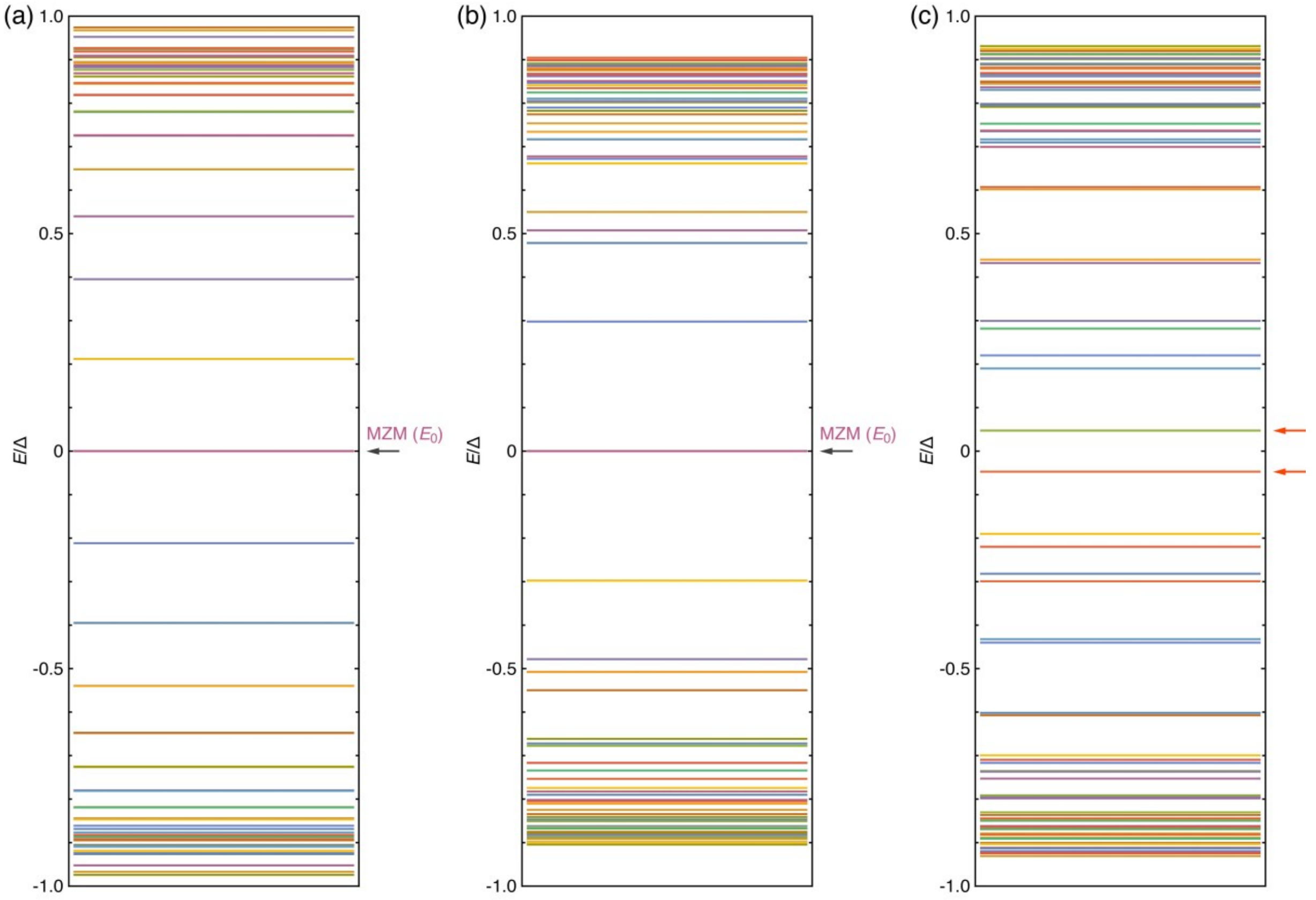

**Fig. S14.** (a) Energy spectrum of a 2D vortex from SC TSS, involving no CDW and $\Delta_Q(x)$. Calculation parameters: Fermi velocity $v_F = 3$, uniform superconducting gap $\Delta_0 = 0.3$, coherence length $\xi = 10$, and Fermi energy $E_F = 1.2$. These parameter values are consistently used in (b) and (c). The spectrum exhibits quantized $E_\mu$ and a clear zero-energy MZM (indicated by the black arrow). (b) Energy spectrum in the presence of $\Delta_Q(x)$ with $d = 0$. The parameters are: amplitude $\Delta_P = 0.15$, wave vector $Q = 2\pi/15$, and dislocation phase $\varphi(x) = 0$. The MZM remains robust under this dislocation-free configuration (black arrow). (c) Energy spectrum in the presence of $\Delta_Q(x)$ for $d = 1/4\lambda$ and a double dislocation phase $\varphi(x) = \pi\Theta(-x)$. The double dislocation disrupts the topological protection of the MZM, resulting in the destruction of the zero-energy state (red arrows).